\magnification1200

\tolerance=2400

\baselineskip14pt

 at 10truept

\font\reg=cmr10 at 12truept

\font\large= cmr16 at 16truept

\font\llarge=cmr20 at 20truept

\font\lllarge=cmr24 at 24truept

{} \vskip 5cm

\lllarge

\centerline{ Perturbation Theory from Automorphic Forms}

\vskip 2cm

\llarge \centerline{Neil Lambert  and Peter West }

\bigskip
\large \centerline{Department of Mathematics}

\centerline{King's College, London}

\centerline{ WC2R 2LS, UK}

\bigskip
\reg\centerline{neil.lambert@kcl.ac.uk, peter.west@kcl.ac.uk}

\bigskip
\large \centerline{\sl Abstract}

\reg Using our previous construction of Eisenstein-like automorphic
forms we derive formulae for the perturbative and non-perturbative
parts  for any group and representation. The result is written in
terms of the weights of the representation and the derivation is
largely group theoretical. Specialising to the $E_{n+1}$ groups
relevant to type II string theory and the representation associated
with node $n+1$ of the $E_{n+1}$ Dynkin diagram we explicitly find
the perturbative part in terms of String Theory variables, such as
the string coupling $g_d$ and volume $V_n$. For dimensions seven and
higher we find that the perturbation theory involves only two terms.
In six dimensions we construct the $SO(5,5)$ automorphic form using
the vector representation. Although these automorphic forms are
generally compatible with String Theory, the one relevant to $R^4$
involves terms with $g_d^{-6}$ and so is problematic. We then study
a constrained $SO(5,5)$ automorphic form, obtained by summing over
null vectors, and compute its perturbative part. We find that it is
consistent with String Theory and makes precise predictions for the
perturbative results. We also study the unconstrained automorphic
forms for $E_6$ in the $\bf {27}$ representation and $E_7$ in the
$\bf 133$ representation, giving their perturbative part and
commenting on their role in String Theory.

\bigskip
\reg \noindent

\vfill \eject

{\large {1. Introduction}}
\bigskip
The low energy effective actions for the type II superstring
theories are the IIA  [1,2,3] and IIB [4,5,6] supergravity theories.
They are complete in that they contain all perturbative and
non-perturbative effects and are essentially unique. To date there
does not exist a non-peturbative formulation of String Theory and
these supergravity theories, as well as the supergravity theory in
eleven dimensions [7], have been essential for our present
understanding.

The maximal supergravity theories in dimensions less than ten can be
obtained from the eleven or ten dimensional theories by dimensional
reduction. One of the most unexpected developments was the
realisation that these theories possess exceptional symmetries. If
one dimensionally reduced the IIB theory in ten dimensions on an $n$
torus then the resulting theory has an $E_{n+1}$ symmetry. In
particular one finds an $E_6$ symmetry in five dimensions, $E_7$
symmetry in four dimensions [8], $E_8$ in three dimensions [9] and
$E_9$ in two dimensions [10,11,12]. In six, seven  and eight
dimensions we find the groups $E_5=SO(5,5)$, $E_4=SL(5)$ and
$E_3=SL(2)\times SL(3)$ respectively [13]. While the IIB
supergravity theory in ten dimensions has an $E_1=SL(2)$ symmetry
[4].

All these supergravity theories possess solitonic solutions whose
charges obey quantisation conditions [14,15] and as the $E_{n+1}$
symmetry groups act on the charges only a discrete subgroup can
persist in the quantum theory, instead of the continuous symmetries
found in the supergravity theories. Such phenomena were first
pointed out in a supergravity context for the four-dimensional
heterotic string [16], which had already been conjectured to possess
an $S$ duality [17]. For type II String Theories  in all dimensions
it was conjectured that a discrete $E_{n+1}$ symmetry, well-known as
U-duality, were symmetries of the full quantum theory [18].

Despite the intense interest in U-duality there has only been
limited work testing this conjecture beyond the low energy, {\it
i.e.} supergravity, approximation. The most notable exception has
been the ten dimensional IIB string theory. In particular, it was
proposed that the coefficients of the $R^4$ and $D^2R^4$ and certain
other higher order terms were specific non-holomorphic automorphic
forms for $SL(2,Z)$ of Eisenstein type [19-25]. These objects were
analysed and it was found that they predicted all the perturbative
and non-perturbative behaviour associated with these corrections.
Indeed these authors were able to show that there was considerable
agreement between these predictions at the perturbative and
non-perturbative level and the ones of IIB string theory. Terms of
the form $R^4H^{4g-4}$ were also considered and found to have
similar interesting features [26].

The evidence for U-duality  in lower dimensions is less strong.
Automorphic forms  in eight dimensions for the group $SL(2)\otimes
SL(3)$ [27,28], and also for seven dimensions for the group $SL(5)$
[27],  have been considered and some agreement with string theory
predictions has been found.  A general discussion of automorphic
forms for $SL(N)$ and  $SO(d,d)$ was given in reference [29] which
derived some of their properties and considered their connection to
string theory, in particular the relation between BPS states and
constrained automorphic forms.

A slightly different approach was taken in references [30-33] which
directly looked for evidence for $E_{n+1}$ symmetries in the higher
derivative corrections. One of the simplest ways to find strong
evidence for the existence of the $E_{n+1}$ symmetries in the
supergravity theories is to compute the dependence of the theory on
the diagonal components of the metric $\vec \phi$ associated with
the $n$-torus. One finds  terms with factors of $e^{\sqrt 2 \vec
w\cdot \phi}$ where the vectors $\vec w$ are the roots of $E_{n+1}$.
Thus although not a proof, one clearly sees the $E_{n+1}$ Lie
algebra emerge in a very transparent way. The same calculation for
the higher derivative corrections does not lead to roots but rather
the weights of $E_{n+1}$ [30,31]. Reference [31] also gave   a
general construction of non-holomorphic  automorphic forms for any
group $G$, based on the theory on non-linear realisations, including
those that transformed non-trivially. An automorphic form was
constructed from a given representation of $G$ and it was shown that
it involved weights of $G$.  Thus the appearance of weights of
$E_{n+1}$ in the dimensional reduction was evidence for the
appearance of non-holomorphic automorphic forms. Furthermore the
highest weight of the representation  on which the automorphic form
was based could be deduced from the dimensional reduction.  Thus the
dimensional reduction of the higher derivative corrections provides
evidence for the appearance of automorphic forms and so for an
$E_{n+1}$ symmetry in the higher derivative corrections [31].

Demanding a discrete $E_{n+1}$ invariance of the complete string
theory effective action  implies that the coefficients of the higher
derivative terms transform in a specific way, which is the same as
the automorphic forms that are well studied in the mathematical
literature. The class of functions  described  in   reference [31],
and  which are used in this paper, are constructed to have the
correct transformation property suitable for any higher derivative
term.  However, it was not demanded that   these functions should be
eigenfunctions of the Laplacian or other Casimir operators, unlike
the automorphic forms in the mathematical literature. While it is
known that the coefficients of certain higher derivative terms that
have low numbers of spacetime derivatives are eigenfunctions of the
Laplacian [21] this is not the case in general ({\it e.g.} see
[23]). Therefore we will use the term automorphic form to be any
function with the appropriate transformation property under
$E_{n+1}$ and not impose any constraints on Casimir operators.

We will also   consider automorphic forms constructed as in [31] but
whose lattice sums are subject to $E_{n+1}$-invariant constraints.
In certain cases imposing quadratic constraints does result in an
eigenfunction of the Laplacian [29]. However, as already  pointed
out in  [31], the automorphic forms that occur, at least for low
numbers of spacetime   derivatives, are likely to also be
eigenfunctions of  the higher order Casimir operators, and these can
presumably be obtained by imposing higher order $E_{n+1}$-invariant
constraints.

It has become clear that higher derivative string corrections in
$d=10-n$ dimensions are controlled by non-holomorphic automorphic
forms of $E_{n+1}$ and these contain all the perturbative and
non-perturbative effects. One of the most striking features to
emerge was that there were very few perturbative corrections
implying novel non-renormalisaton theorems, some which have been
verified in String Theory [34] and state that certain operators only
receive contributions from two orders of perturbation theory.
However, the relatively small number of papers on this  subject is a
testament to the difficulties of working with non-holomorphic
automorphic forms, some recent studies include [35],[36].

This paper is organised as follows. In section 2 we will use our
previous method [31] to construct Eisenstein-like automorphic forms
for any group representation and in section 3  find explicit
formulae for the ``perturbative" and ``non-perturbative" parts.
Since our construction is based on representation theory the
derivation of these formulae is largely group theoretic and the
result is expressed in terms of the weights of the representation
being used to construct the automorphic form. In section 4 we will
discuss the analytic continuation and regulation of these forms that
is required to define the automorphic forms that are expected to be
relevant in String Theory. In section 6 we then apply the
perturbative formula to the groups $E_{n+1}$ and the fundamental
representation associated with node ${n+1}$ of the $E_{n+1}$ Dynkin
diagram (see figure 1), finding an explicit form of the perturbative
part using the decomposition to $SL(2)\otimes SL(n)$. In section 7
we will compute the perturbative part in terms of the string
quantities, that is the string coupling $g_d$ in $d$ dimensions and
the volume of the torus $V_n$.

We will find that for the case of dimensions $d\ge 7$, that is automorphic forms for the groups $SL(2)$,
$SL(2)\otimes SL(3) $ and $SL(5) $, that there is   a physically acceptable perturbative series with
contributions at only two orders. Acceptable here simply means it is consists of terms of the form
$g_d^{2g-2}$. This a necessary condition for the automorphic form to appear in String Theory but it is
certainly not a sufficient condition. However, this is not always the case for six dimensions, that is for the
automorphic forms based on the group $SO(5,5)$ and the vector, {\it i.e.} the ${\bf 10}$, representation. In
dimensions five and four we consider the automorphic forms for the exceptional groups $E_6$, with
representation $\bf 27$, and $E_7$, with representation $\bf 133$, respectively. We discuss to what extend
these perturbation series are consistent with string theory. In section 8 we then consider constrained
$SO(5,5)$ automorphic forms based on a null vector and find that one always has an acceptable perturbation
theory result which is explicitly derived in terms of string theory variables. Section 9 contains a discussion
of our results.

\bigskip
{\bf Note Added}: While this paper was in preparation we received
[37] which contains some overlap of our work, in particular section
8. In addition  reference [29] was recently revised and shortly
after this paper appear we received [38] which also contains related
results.

\bigskip
{\large {2. Construction of Automorphic Forms for $G/H$}}
\bigskip

Let us begin by a review of the  construction of non-linear realisations. We consider a group $G$ with Lie
algebra $Lie(G)$. $Lie(G)$ can be split into the Cartan subalgebra with elements $\vec H$, positive root
generators $E_{\vec\alpha}$ and negative root generators $E_{-\vec \alpha}$ with $\vec\alpha>0$. There exists a
natural involution, known as the Cartan involution, defined by
$$
\tau :(\vec H,E_{\vec\alpha})\to -(\vec H, E_{-\vec\alpha})\ , \eqno (2.1)
$$
which can extended to the group by defining $\tau(g_1g_2)=\tau(g_1)\tau(g_2)$. To construct the  non-linear
realisation we must specify a subgroup $H $ (not to be confused with the generators of the Cartan subgroup
which are denoted by $\vec H$). For us this is defined to be  the subgroup left invariant under the Cartan
involution, {\it i.e.} $H=\{g\ \in \ G:\tau(g)=g\}\equiv I(G)$. In terms of the Lie algebra $Lie(I(G))$ it is
all elements $A$ such that $A= \tau(A)$.

The non-linear realisation is constructed from group elements $g(x) \in G$ which in physical applications
depend on the spacetime coordinates $x^\mu$. These are subject to the transformations
$$
g(x)\to g_0 g(x)h^{-1}(x) \eqno (2.2)
$$
where $h(x)\in H$ and also depends on spacetime. We may write the group element in the form $ g(x) = e^
{\sum_{\vec\alpha>0} \chi_{\vec\alpha} E_{\vec\alpha}} e^{-{1\over \sqrt{2}}\vec\phi\cdot\vec H} e^{\sum_
{\vec\alpha>0} u_{\vec\alpha} E_{-\vec\alpha}}$, but using the local transformation we can bring it to the form
$$
g(\xi) = e^{\sum_{\vec\alpha>0} \chi_{\vec\alpha} E_{\vec\alpha}} e^{- {1\over \sqrt{2}}\vec\phi\cdot\vec H}\
.\eqno(2.3)
$$
Here we use  $\xi = (\vec\phi,\chi_{\vec\alpha})$ as a generic symbol for all the scalar fields, which are
functions of spacetime,  that parameterize the coset representative. Under a rigid $g_0\in G$ transformation
$g(\xi) \to g_0 g(\xi)$ this form for the coset representative is not preserved. However one can make a
compensating transformation $h(g_0,\xi)\in H$ that returns $g_0g(\xi) $ into the form of equation (2.3);
$$
g_0g(\xi)h^{-1}(g_0,\xi) = g(g_0\cdot \xi)\ .\eqno (2.4)
$$
This induces a non-linear action of  the group $G$ on the scalars; $\xi \to g_0\cdot \xi$.

Using the Cartan involution we can define the notion of a `generalized transpose';
$$
A^\#= -\tau(A)\ .\eqno(2.5)
$$
which in terms of group elements is given by  $g^\#= (\tau (g))^{-1} $. We note that for group elements $h\in
H$   $h^{\#}=h^{-1}$. Unlike $\tau$, the operation $\#$ inverts the order of two group elements {\it i.e.}
$(g_1g_2)^\# =g_2^\#g_1^\#$ and also on the product of two elements of the algebra.

Let us now review the construction automorphic forms given in [31]. We will need   a  linear representation of
$G$. Let $\vec\mu^i$, $i=1,...,N$ be the weights   of the representation and $|\vec\mu^i>$ be a corresponding
states. The weights of the representation can be ordered by saying $\vec\mu^i>\vec \mu^j$ iff
$\vec\mu^i-\vec\mu^j=\vec\alpha_{ij}$ is a positive element of the root lattice (note that this requires that
one chooses an ordering of the roots). If there are non-zero multiplicities then one can choose any ordering of
the degenerate weights that one likes. We will choose to order the weights such that
$\vec\mu^1>\vec\mu^2...>\vec \mu^N$. Thus $\vec\mu^1$ is the highest weight. The corresponding state satisfies
$E_{\vec\alpha}|\vec\mu^1>=0$ for all simple roots $\vec\alpha$ the states in the rest of the representation
are polynomials of $F_{\vec \alpha}= E_{-\vec\alpha}$ acting on the highest weight state.

We consider states of the form $|\psi>=\sum_i\psi_i |\vec\mu^i>$. Under the action $U(g_0)$ of the group $G$ we
have
$$
|\psi>\to U(g_0)|\psi>=L( g_0^{-1}) \sum_i \psi_i|\vec\mu^i> \equiv (U (g_0)\psi_i) |\vec\mu^i>= \sum_{i,j}
D_i{}^j (g_0^{-1})\psi_j |\vec\mu^i> \eqno (2.6)
$$
where $L(g_0) $ is the expression of the group element $g_0$ in terms of the Lie algebra elements which now act
of the states of the representation in the usual way. We note that the action of the group on the components
$\psi_i$ is given by $ \psi_i\to U(g_0)\psi_i= \sum_j D_i{}^j (g_0^ {-1})\psi_j $ which is the result expected
for a passive action. From this equation we can use the action of the Lie algebra elements on the states of the
representation to compute the matrix $ D_i{}^j$ of the representation.

Given any linear representation $\psi$, we can construct the Cartan involution twisted representation, denoted
by $\psi_\tau$, which by definition transforms as
$$
|\psi_\tau>\to U(g_0)|\psi_\tau>=L( g_0^\#) |\psi_\tau> \ ,\eqno (2.7)
$$
We will also need  the dual representation, denoted by $<\psi_D |$, which transforms as
$$
<\psi_D | \to U(g_0) (<\psi_D | )= <\psi_D | L(g_0) \ ,\eqno (2.8)
$$
It is constructed just so that $ <\psi_D |\psi>$ is invariant. Using
both constructions we have the dual twisted representation
$<\psi_{D\tau} |$ which transforms as $<\psi_ {D \tau} |\to
U(g_0)<\psi_{D\tau} |= <\psi_{D\tau} |L(\tau(g_0))$. The
representation $<\psi_{D\tau} |$ has the same highest weight as
$|\psi>$ and so we can identify it as the same representation [31].

Given any linear realisation, such as the one in equation (2.6), we can construct a non-linear realisation by
$$\eqalign{
|\varphi(\xi)> &= \sum\varphi_i(\xi) |\vec\mu^i> \cr &=L((g(\xi))^ {-1})|\psi>\cr &= e^{\sum_{\vec\alpha>0}
e^{{1\over \sqrt{2}}\vec\phi\cdot\vec H}}e^{-\sum_{\vec\alpha>0}\chi_ {\vec \alpha} E_{\vec\alpha}} |\psi> }\
,\eqno (2.9)
$$
where $g(\xi)$ is the group element of the non-linear realisation in equation (2.3). Under a group
transformation $U(g_0)$ it transforms as
$$\eqalign{
U(g_0)|\varphi(\xi)> &= L((g(\xi))^{-1})U(g_0)|\psi> \cr &=L((g(\xi))^ {-1}) L(g_0^{-1} ) |\psi> \cr &= L(
(g_0g(\xi))^{-1})|\psi>\cr &= L(h^{-1})|\varphi(g_0\cdot\xi)> \ ,\cr} \eqno (2.10)
$$
using equation (2.4). In terms of the component fields we find that $ \varphi_i (\xi)= \sum_j
D_i{}^j((g(\xi))^{-1})\psi_j$ and $U(g_0)\varphi_i (\xi)= \sum_j D_i {} ^j((h)^{-1})\varphi_j (g_0\cdot \xi)$.
The reader can find the examples for $SL(2)$ and $SL(3)$ worked out in equations (5.9), (5.15) and (5.22)
respectively.

For the dual and Cartan twisted representations of equations (2.9) and (2.10) we can also introduce
non-linearly transforming representations. Indeed for the case of the dual twisted representation we find
$$
<\varphi_{D\tau}(\xi) |= <\psi_{D\tau} |L(\tau(g(\xi))) \ ,\eqno (2.11)
$$
which transforms as $U(g_0) <\varphi_{D\tau}(\xi) | = <\varphi_{D \tau}(g_0\cdot \xi)| L(h)$. We note that
$$
<\varphi_{D\tau}(\xi) |\varphi (\xi)>\to U(g_0)(<\varphi_{D\tau}(\xi) |\varphi (\xi)>)= (<
\varphi_{D\tau}(g_0\cdot \xi) |\varphi (g_0\cdot \xi)>) \ ,\eqno (2.12)
$$

An automorphic form $\Phi(\xi)$ is a function on $G/H$ that satisfies
$$
\Phi(g_0\cdot \xi) = D(h(g_0,\xi)^{-1})\Phi(\xi)\ ,\eqno (2.13)
$$
for some representation $D$ of $H$ where now the group $G$ is now taken to be a discrete subgroup of $G$. In
this paper we will take the trivial representation and hence have
$$
\Phi(g_0\cdot\xi) = \Phi(\xi)\ .\eqno (2.14)
$$
Given a linear representation with highest weight $|\vec \mu^1> $ we consider states of the form
$$
|\psi>= \sum_im_i|\vec\mu^i>,\ \  m_i\in {\bf Z}\ . \eqno (2.15)
$$
Thus our components are now the integers $m_i$. We can think of these states as belonging to a lattice
$$
\Lambda= \{\sum_im_i|\vec\mu^i>| \  m_i\in {\bf Z}\ {\rm not\ all\ vanishing}\}, \eqno (2.16)
$$
with the origin deleted. The discrete version of $G$ that we obtain then consists of elements of $G$ which
preserve the lattice. The precise details of this can be rather subtle and we will not comment more on it here.

We then construct the non-linear realisations $\varphi (\xi)>$  using the states of equation (2.13) in
equations (2.9) and similarly and $\varphi_{D\tau} (\xi)>$ using equation (2.11);
$\phi_{D\tau}=\sum_j<\mu^j|m_j$.The invariant  automorphic form is given by
$$
\Phi(\xi) = \sum_\Lambda
F(u(\xi))\ .\eqno (2.17)
$$
where the sum is over the integers $m_i$ that occur in the lattice $ \Lambda$, $F$
is a function of $u$ and
$$\eqalign{
u(\xi)&= <\varphi_{ D\tau }(\xi) |\varphi(\xi) > \cr & =<\psi_{ D\tau}|L(\tau (g(\xi)))
L((g(\xi))^{-1})|\psi>\cr &=\sum_{i,j}<\vec \mu^j|m_jL(\tau(g(\xi))) L((g(\xi))^{-1}) m_i |\vec \mu^i>\cr &=
\sum_{i,j}<\vec\mu^j|m_j( e^{-\sum_{\vec\alpha>0}E_{-\alpha }\chi_ \alpha}  e^{ \sqrt{2}\vec\phi\cdot\vec H}
e^{-\sum_{\vec\alpha>0}E_\alpha \chi_\alpha}) m_i |\vec\mu^i> \ .\cr} \eqno (2.18)
$$
Under a group $G$ transformation
$$
u(\xi)\to U(g_0)u(\xi)= u(g_0\cdot \xi^\prime) \ . \eqno (2.19)
$$
However, in the automorphic form of equation (2.17) a $U(g_0)$
transformation on $|\psi >$ is just a rearrangement of the integers
$m_i$ and this can be undone by a change  of summation of over the
lattice. As a result we find that the automorphic form of equation
(2.17) transforms as in equation (2.14) as required. The particular
case that $F(u) = u^{-s}$ the corresponding automorphic form will
simply be denoted by $\Phi_s$. We note that there are additional
possible definitions of Eisenstein-like series that we will not
discuss here, for example one can construct automorphic forms where
where each fundamental weight has a complex number $s$ associated to
it. Such forms were recently considered in a String Theory context
in [36].

We can construct another automorphic form using  $|\psi_{\tau}>$ and $< \psi_{D}|$. Let us define
$$
v (\xi)= <\varphi_{ D}(\xi) |\varphi_{\tau}(\xi) > =<\psi_{ D}|L((g (\xi))) L((g(\xi))^\#)|\psi_\tau> \ .\eqno
(2.20)
$$
Using similar arguments once can show that $v(\xi)\to U(g_0)v(\xi)= v (g_0\cdot \xi^\prime)$ and that
$$
\Phi^{\tau (R)} (\xi)= \sum _\Lambda F(v) \ .\eqno (2.21)
$$
is an automorphic form. We will next show that it is related to that of equation (2.17). If we take $F(v)=
v^{-s}$ we denote the automorphic form by $\Phi^{\tau (R)}_s (\xi)$.

We began with the automorphic form constructed from the representation $R$ and write it as
$$
{\Gamma(s)\over \pi^s}\Phi_s=\sum_{m_i}  \int {dt \over t^{1+s}}e^{-{\pi\over t} m_i A^{ij} m_j }\ ,\eqno(2.22)
$$
where $A=(D((g(\tau (g))^{-1})^{-1})^{ij}=D(\tau(g)g^{-1})^{ij}$. Poisson resuming, using equation (A.2),  we
find that
$$
{\Gamma(s)\over \pi^s}\Phi_s = \sum_{\hat m^i}  (det A)^{-{1\over 2}} \int {dt \over t^{1+(s-N/2)}}e^{-\pi t
\hat m^i (A^{-1})_{ij}\hat m^j} \eqno(2.23)
$$
We note that $A^{-1}= D(g\tau(g^{-1}))_{ij}$.  Changing integration variable $t$ to $1/t$ gives
$$\eqalign{
{\Gamma(s)\over \pi^s}\Phi_s &=  {\Gamma({N\over 2}-s)\over \pi^{{N\over2}-s}} \sum_{\hat m^i} (det
A)^{-{1\over 2}} \int {dt \over t^{1-s+N/2}}e^{-{\pi\over t} \hat m^i (A^{-1}))_{ij}\hat m^j}\cr &=
{\Gamma({N\over 2}-s)\over \pi^{{N\over 2}-s}}(det A)^{-{1\over 2}}\sum_{\hat m^i}{1\over (\hat m^i
(A^{-1}))_{ij}\hat m^j)^{{N\over 2}-s}} \cr}\eqno(2.24)
$$
We can interpret this as the  automorphic form constructed from the
representation $\tau(R)$ as $g\to \tau (g)$ takes $A\to A^{-1}$. We
note that if the representation $R$ has highest weight $\vec\mu^1$
and lowest weight $\vec \mu^N$ then the Cartan involution  twisted
representation $\tau ( R)$ has highest weight $-\vec\mu^N$ and
lowest weight $-\vec \mu^1$. In fact these representations  are
related by $-W_0$, where $W_0$ is the unique Weyl reflection with
the longest length (see appendix B of reference [31]). The latter is
an automorphism of the Dynkin diagram and hence also the Lie algebra
and therefore these two representations are related by this
automorphism. To give an example,  if the representation $R$ of
$SL(N)$ is the $N$ representation then the representation $\tau (R)$
is the $\bar N$ representation. The automorphism that relates these
representations is the one that exchanges the nodes $i\to N-i$ of
the $SL(N)$ Dynkin diagram. In this case swapping the fundamental
representation associated with node one to that associated with node
$N-1$.

In general the sum over the integers $\hat m^i$ can be interpreted as the lattice associated with the
representation $\tau (R)$ of the group $G $. As such we have demonstrated that
$$ {\Gamma(s)\over \det (D (\tau (g(\xi)))\pi^s}\Phi_s^R
={\Gamma(N/2-s) \over \det (D( g(\xi)) \pi^{N/2-s}}\Phi^{\tau (R)}_{N/2-s} \ .\eqno(2.25)
$$
The automorphic form that appears on the left hand side of this relation is just that of equation (2.17), with
$F(u)={1\over u^s}$ and it is built using the representation $R$, although it was denoted by just $\Phi_s$
there,   we above have denoted it by $\Phi_s^R$. The automorphic forms on the right hand side is of the form of
that  in equation (2.21) and it is built using the representation $\tau (R)$ and so we denoted it by
$\Phi^{\tau (R)}_s$ if $F(v)={1\over v^s}$.

The alert reader will have noticed that we have  added and then subtracted the divergent terms at $m_i=0$ and
at $\hat m^i=0$ respectively when carrying out the Poisson resummation. As such we have assumed that these
terms can be regulated and they do not change the result.

\bigskip
{\large {3. Evaluation of Automorphic Forms}}
\bigskip

In this section which wish to develop explicit expressions for the Einstein-like series of Automorphic forms
defined in the previous section by
$$
\Phi(\xi) = \sum_\Lambda{1\over (u(\xi))^s}\ . \eqno(3.1)
$$
where $u$ is given in equation (2.18). This sum is convergent and $\Phi$ is well-defined whenever $s>N/2$ as
can be seen heuristically by taking the sum to be an integral. However, it can be defined by analytic
continuation for almost all other values of $s$. As explained in section two we write the states of the
representation $R$ as $|\psi> = \sum_i m_i|\vec\mu^i>$ so that the state of equation (2.9)
$$
|\varphi> = \sum_i m_i e^{{1\over \sqrt{2}}\vec\phi\cdot\vec H} e^{-\sum_{\vec\alpha}\chi_{\vec\alpha}
E_{\vec\alpha}}|\vec\mu^i>\ .\eqno(3.2)
$$
Since we know the action of the Lie algebra generators on the weights we can evaluate
$L(e^{-\sum_{\vec\alpha}\chi_{\vec\alpha} E_{\vec\alpha}})$ on the representation in a straight forward way
once we take
$$
L(E_{\vec\alpha})|\vec\mu^i> = c_{\vec\alpha i}|\vec\mu^i+\vec\alpha>\ , \eqno(3.3)
$$
for some constant $c_{\vec\alpha i}$ whose values  we will comment later. As a result we find that we can write
$$
L(e^{-\sum_{\vec\alpha}\chi_{\vec\alpha} E_{\vec\alpha}})|\vec\mu^i> = |\vec\mu^i> -
\sum_{j}\tilde\chi_{ji}|\vec\mu^j>\ .\eqno(3.4)
$$
which defines the symbols $\tilde\chi_{ji}$. Taking the innerproduct with $\omega_k^{-1}|\vec\mu^k>$ we find
$$
\eqalign{\tilde\chi_{ki} &= \delta_{ki}  - \omega_k^{-1}<\vec\mu^k|L(e^{-\sum_{\vec\alpha}\chi_{\vec\alpha}
E_{\vec\alpha}})|\vec\mu^i>\cr &=-\sum_{n=1}^\infty {(-1)^n\over n! \omega_k}\sum_{\vec\alpha_1}\ldots
\sum_{\vec\alpha_n}\chi_{\vec\alpha_1}\ldots\chi_{\vec\alpha_n} <\vec\mu^k|L(E_{\vec\alpha_1}\ldots
E_{\vec\alpha_n})|\vec\mu^i>\cr &=c_{\vec\alpha_{ki} i}\chi_{\vec\alpha_{ki}} +
{Poly}(\chi_{\vec\beta},0<\vec\beta< \vec\alpha_{ki})\ ,\cr} \eqno(3.5)
$$
where $\vec\alpha_{ki} = \vec\mu^k-\vec\mu^i$ and $Poly$  is a polynomial in $\chi_{\vec\beta}$ that only
involves roots  such that $0<\vec\beta<\vec\alpha_{ki}$. Note also that $\tilde\chi_{ki}=0$ unless
$\vec\alpha_{ki}=\vec\mu^k-\vec\mu^i$  is a positive element of the root lattice, {\it i.e. if $k<i$}. We have
normalised the necessarily orthogonal states of the representation as
$<\vec\mu^k|\vec\mu^i>=\omega_i\delta_{ik}$

Thus  the state of equation (3.2) is  then given by
$$
\eqalign{|\varphi> & = \sum_im_i L(e^{{1\over \sqrt{2}}\vec\phi\cdot\vec H})\left(|\vec\mu^i> -
\sum_k\tilde\chi_{ki}|\vec\mu^k>\right)\cr &= \sum_im_i\left( e^{{1 \over \sqrt{2}}\vec\phi\cdot\vec
\mu^i}|\vec\mu^i> - \sum_k\tilde\chi_{ki} e^{{1\over \sqrt{2}}\vec\phi \cdot\vec \mu^k}|\vec\mu^k>\right)}\ .
\eqno(3.6)
$$
Rearranging the sum over $i$ this can be rewritten as
$$
|\varphi> = \sum_i \left(m_i -\sum_{j>i}\tilde\chi_{{ij}}m_j\right)e^ {{1\over \sqrt{2}}\vec\phi\cdot\vec
\mu^i}|\vec\mu^i>\ .\eqno(3.7)
$$
Thus we find the $u$ of the automorphic form of equation (2.18) is given
$$
u = \sum_{i=1}^N (m_i-\tilde \chi_i)^2\omega_ie^{\sqrt{2}\vec\phi\cdot\vec\mu^i}\ ,\eqno(3.8)
$$
where
$$
\tilde \chi_i  = \sum_{j>i}\tilde\chi_{ij}m^j\ .\eqno(3.9)
$$
Note that all the dependence of  the axion-like fields $\chi_{\vec\alpha}$ are contained in the $\tilde
\chi_i$. Since $\tilde \chi_{ij}=0$ if $i> j$ we find that $\tilde \chi_j$ depends only on $m_k,\ k>j$.  In
particular the only dependence on $m_1$ in $u$ is  the explicit $m_1$ in the first term.

In the above we have introduced a normalisation of the states $\omega_i$ and a constant $c_{\vec\alpha i}$ in
the action of the $E_\alpha$ generators of equation (3.3). Clearly, we are free to choose one or the other of
these constants by scaling the states of the representation. In particular if we choose $\omega_i=1$ then we
can set $L(E_{\vec\alpha_a})|\vec\mu^i> =N_{\alpha_a}|\vec\mu^i+\vec\alpha_a> $ for any simple root $\alpha_a$
and then determine all the constants $N_{\alpha_a}$ by implementing the Lie algebra relations. For example if
one does this for $SL(N)$ one finds that all the  $N_{\alpha_a}=0,1$ but this will not be the case in general.

It will prove useful to write, using equation (A.1),  the automorphic form as
$$
\Phi = \sum_\Lambda{1\over u^s} =\sum_\Lambda {\pi^s\over\Gamma(s)}\int_0^\infty {dt\over t^{1+s}}e^{-{\pi\over
t}u}\ ,\eqno(3.10)
$$
We can split the sum into two pieces;  a first terms that is over only $m_1\ne 0$ with all other $m_j=0$ for
$j>1$ and a second term that is over over all $m_1$, including zero, and over all $m_j,\ j>1$, but not
including $m_j=0,\ j=2,3,\dots $. The result is
$$\eqalign{
\Phi_s^N &= \sum_{m_1\ne 0} {\pi^s\over\Gamma(s)}\int_0^\infty {dt\over t^{1+s}}e^{-{\pi\over t}
m_1^2\omega_1e^{\sqrt{2} \vec\phi\cdot\vec\mu^1}}\cr &+ \sum_{ m_1=-\infty}^\infty\sum_{\Lambda_{1}}
{\pi^s\over\Gamma(s)}\int_0^\infty {dt\over t^{1+s}}e^{-{\pi\over t} (m_1-\tilde
\chi_1)^2\omega_1e^{\sqrt{2}\vec\phi\cdot\vec\mu^1}} e^{-{\pi\over t}{u_1}}\ ,\cr} \eqno(3.12)
$$
where
$$
u_1 = \sum_{i>1} (m_i-\tilde \chi_i)^2{\omega_i} e^{\sqrt{2}\vec\phi\cdot\vec\mu^i}\ .\eqno(3.13)
$$
and $\Lambda_{1}$ is the lattice spanned by  $m_2,...,m_{N}\in {\bf Z}$, with the origin omitted. We note that
there is no $\tilde \chi_1$ in the first term since  it depends on $m_j,\ j>1$, but in this term we have
$m_j=0$ for $j>1$. We have dented  $\Phi$ by $\Phi^N_s$ for reasons that will become apparent; the $N$ refers
to the dimension of the lattice.

The first term is simply evaluated using equation (A.1). While the sum in the second  term is over all integers
$m_1$ and so we may use Poisson resummation, that is  equation (A.2). Carrying this out we find that
$$\eqalign{
\Phi_s^N &= 2\zeta(2s){1\over \omega_1^s} e^{-\sqrt{2}s\vec\phi\cdot\vec\mu^1}\cr &+ \sum_{\hat
m_1=-\infty}^\infty\sum_{\Lambda_{1}} {\pi^s\over\Gamma(s)}\int_0^\infty {dt\over
t^{1+s}}\sqrt{{t\over\omega_1}} e^{-{1\over\sqrt{2}}\vec\phi\cdot\vec\mu^1}e^{2\pi i\hat m_1\tilde \chi_1}\cr
&\hskip5cm\times e^{-\pi t \hat m_1^2\omega_1^{-1}e^{-\sqrt{2}\vec\phi\cdot\vec\mu^1}} e^{-{\pi\over t}{u_1}}\
.\cr} \eqno(3.14)
$$

We will now construct a recursion relation in $N$ and $s$ by   splitting  the second term into a piece
corresponding to $\hat m_1=0$ and the rest
$$\eqalign{
&\Phi_s^N = 2\zeta(2s){1\over \omega_1^s} e^{-\sqrt{2}s\vec\phi\cdot\vec\mu^1}\cr &+
\sum_{\Lambda_{1}}\sqrt{{\pi\over\omega_1}} {\Gamma(s-1/2)\over\Gamma(s)}e^{-{1\over\sqrt{2}}
\vec\phi\cdot\vec\mu^1}{\pi^{s-1/2}\over\Gamma(s-1/2)}\int_0^\infty {dt\over t^{1+(s-1/2)}}e^{-{\pi\over
t}u_1}\cr &+ {2\over\sqrt{\omega_1}}\sum_{\hat m_1=1}^\infty\sum_{\Lambda_{1}}\cos(2\pi \hat m_1\tilde \chi_1)
{\pi^s\over\Gamma(s)}e^{-{1\over\sqrt{2}}\vec\phi\cdot\vec\mu^1} \int_0^\infty {dt\over t^{1+(s-1/2)}}e^{-\pi t
\hat m_1^2\omega_1^{-1}e^{-\sqrt{2}\vec\phi\cdot\vec\mu^1}}e^{-{\pi \over t}u_1}\cr &=2\zeta(2s){1\over
\omega_1^s}e^{-\sqrt{2}s\vec\phi\cdot\vec\mu^1}+ \sqrt{{\pi\over \omega_1}}
{\Gamma(s-1/2)\over\Gamma(s)}e^{-{1\over\sqrt{2}}\vec\phi\cdot\vec \mu^1}\Phi^{N-1}_{s-1/2}+ \Upsilon^N_s\ .\cr
}\eqno(3.15)
$$
Here $\Phi^{N-1}_{s-1/2}$ takes the form of the original  function $\Phi^N_s$ but with a shifted valued of $s$
and a lattice $\Lambda_1$ which is  constructed only from the basis of $N-1$ states
$|\vec\mu^2>,...,|\vec\mu^{N}>$. Using the Bessel function integral formula (A.3) we can write $\Upsilon^N_s$
as
$$\eqalign{
\Upsilon^N_s&={4\over(\sqrt{\omega_1})^{s+1/2} }{\pi^s\over\Gamma(s)}
e^{-{1\over\sqrt{2}}(s+1/2)\vec\phi\cdot\vec \mu^1} \sum_{\hat m_1=1}^\infty\sum_{\Lambda_{1}} \left( {\hat
m_1\over \sqrt{u_1} }\right)^{s-1/2}\cos(2\pi \hat m_1\tilde \chi_1)\cr &\hskip6cm \times K_{s-1/2}
\left({2\pi\over \sqrt{\omega_1}}e^{-{1\over\sqrt{2}}\vec\phi\cdot\vec\mu^1} \hat m_1\sqrt{u_1}\right)\
.\cr}\eqno(3.17)
$$
In the limit of large $x$, $K_{s-1/2}(x)\sim e^{-x}$ is  exponentially small. Thus $\Upsilon^N_s$ is
exponentially suppressed in some region of moduli space and we refer to it as  non-perturbative. Substituting
into equation (3.15) we now find the recursion relation for the remaining, perturbative, piece
$$
\Phi_s^N{}_{p}= 2\zeta(2s){1\over \omega_1^s} e^{-\sqrt{2}s\vec\phi\cdot\vec\mu^1}+ \sqrt{{\pi\over\omega_1}}
{\Gamma(s-1/2)\over\Gamma(s)} e^{-{1\over\sqrt{2}}\vec\phi\cdot\vec\mu^1}\Phi^{N-1}_{s-1/2}{}_p . \eqno(3.18)
$$
We can iterate this recursion relation $N$-times to obtain
$$
\Phi_{p}=\sum_{k=1}^N {2\over \omega_k^{s-{k-1\over 2}}} \zeta(2s-k+1)\pi^{k-1\over 2}{\Gamma(s-{k-1\over
2})\over \Gamma(s)}e^{-\sqrt{2}(s-{k-1\over 2})\vec\phi\cdot\vec\mu^k}\prod_{i<k}\sqrt{1\over
\omega_i}e^{-{1\over\sqrt{2}}\vec\phi\cdot\vec\mu^i}\ .\eqno(3.19)
$$

We also obtain a recursion relation  for the remaining non-perturbative part of $\Phi$ by including the
$\Upsilon$-terms:
$$
\Phi^N_s{}_{np} =\sqrt{{\pi\over\omega_1}} {\Gamma(s-1/2)\over\Gamma(s)}e^{-{1\over\sqrt{2}}\vec\phi\cdot\vec
\mu^1}\Upsilon_{s-1/2}^{N-1}+ \Upsilon^N_s\ .\eqno(3.20)
$$
Iterating this gives
$$
\eqalign{\Phi_{np} &= \sum_{k=1}^{N-1}  {4\over (\sqrt{\omega_k})^{s-{k-2\over 2}}}{\pi^s\over
\Gamma(s)}e^{-{1\over\sqrt{2}} (s-{k-2\over 2})\vec\phi\cdot\vec\mu^k}\prod_{j<k}\sqrt{1\over
\omega_j}e^{-{1\over\sqrt{2}}\vec\phi\cdot\vec\mu^j}\cos(2\pi \hat m_k \tilde \chi_k)\cr &\hskip1cm
\times\sum_{\hat m_k=1}^\infty\sum_{\Lambda_{k}} \left( \hat m_k\over {\sqrt{u_k}}\right)^{s-{k\over
2}}K_{s-{k\over 2}}\left({2\pi\over\sqrt{\omega_k}}e^{-{1\over\sqrt {2}}\vec\phi\cdot\vec\mu^k}\hat
m_k\sqrt{u_k})\right)}\ ,\eqno(3.21)
$$
where $\Lambda_{k}$ is spanned by $|\vec\mu^k>,...,|\vec\mu^{N}>$ and
$$
u_k = \sum_{i>k}(m_i-\tilde\chi_i)^2\omega_ie^{\sqrt{2}\vec\phi\cdot \vec\mu^i}\ .\eqno(3.22)
$$
We see that all the dependence on $\chi_{\vec\alpha}$ is contained in the $\Phi_{np}$.  We have therefore
split $\Phi$ into a perturbative and a non-perturbative piece, $\Phi = \Phi_p+\Phi_{np}$. We will see later
when comparing with string theory that this is indeed the correct split.

Equation (3.19) is the main technical result of this paper. It shows that the automorphic forms constructed
above through equation (3.1) for any $N$-dimensional representation of $G$ always have $N$ contributions to
their perturbative part.  However, as we will see below, this does not in general correspond to $N$ distinct
orders of string perturbation theory. These $N$ terms are given by exponentials of the form $e^{{1\over
\sqrt{2}}\vec w\cdot \vec\phi}$ where $\vec w$ are certain linear combinations of the weights of $G$. Thus
determining the perturbative part is essentially reduced to a linear algebra problem in the weight space of the
representation that is used to construct the automorphic form.

\bigskip
{\large {4. Regularization of Automorphic forms}}
\bigskip

As we mentioned above $\Phi_s$ is well defined when $s>N/2$. However
the most well studied higher derivative corrections correspond to
small values of $s$ such as $s=3/2$ for the $R^4$ term. Thus we wish
to extend the definition of  $\Phi_s$ to more general values of $s$.
One can show that $\Phi_ {np}$ is always convergent and hence well
defined. The divergences show up in $\Phi_p$ because $\zeta(z)$ and
$ \Gamma(z)$ are only convergent for $z>1$ and $z>0$ respectively.
However since these functions can be analytically continued we will
be able regulate $\Phi$ and extend its  definition to any value of
$s$. Regularization has also been discussed for some special cases
in [27].

More specifically by analytic continuation $\zeta(z)$ can be defined for all $z\ne 1$ and $\Gamma(z)$ for all
$z\ne 0,-1,-2,...$. These are simple poles with
$$
\zeta(1+\epsilon) = {1\over \epsilon}\qquad \Gamma(-n+\epsilon) = { (-1)^n\over n!\epsilon}\ ,\eqno(4.1)
$$
and a useful formula is
$$
\zeta(-z) = -2^{-z}\pi^{-z-1}\sin\left({\pi z\over 2}\right)\Gamma(1 +z)\zeta(1+z)\ .\eqno(4.2)
$$
Note that $\zeta(-2n)=0$ and hence $\zeta(-2n)\Gamma(-n)$ is finite for $n=1,2,3...$. Thus for generic values
of $s$ the automorphic forms are well-defined using analytic continuation. However for some values, namely when
$s\le N/2$ with $2s$ a positive integer, we encounter divergences. To regulate the automorphic form we can
deform $s\to s+\epsilon$, remove the $1/\epsilon$ pole and then take the limit $\epsilon \to 0$. This will
preserve the automorphic properties of $\Phi$, provided that the residue of the $1/\epsilon$ pole is a
constant.

First consider the case $s<N/2$ with $2s$ a positive integer. The terms in $\Phi_p$ come with coefficients
$$
\zeta(2s-k+1)\Gamma(s-(k-1)/2)/\Gamma(s)\ .
$$
There are potential divergences when $2s-k+1=1$ or $s-(k-1)/ 2=0,-1,-2,-3,...$. However since
$\zeta(-2n)\Gamma(-n)$ is finite if $n=1,2,3,...$ we see that the only problematic terms in the later case
arise when $s-(k-1)/2=0$. Thus there are two divergent terms, one where $k=2s$ and one with $k=2s+1$. Let us
deform $s\to s+\epsilon$ and look at these two terms. We find
$$
\eqalign{ & 2\zeta(1+2\epsilon){\Gamma(1/2+\epsilon)\over\Gamma(s+\epsilon)}\pi^
{s-1/2}e^{-\sqrt{2}(1/2+\epsilon)\vec\phi\cdot\vec\mu^{2s}}
\prod_{i<2s}e^{-{1\over\sqrt{2}}\vec\phi\cdot\vec\mu^{i}} \cr &\qquad+
2\zeta(2\epsilon){\Gamma(\epsilon)\over\Gamma(s+\epsilon)}\pi^{s}e^{-
\sqrt{2}\epsilon\vec\phi\cdot\vec\mu^{2s+1}}\prod_{i<2s+1}e^{-{1\over \sqrt{2}}\vec\phi\cdot\vec\mu^{i}} \ .
\cr }\eqno(4.3)
$$
Substituting in $\zeta(0)=-1/2$, $\Gamma(1/2)=\sqrt{\pi}$ and extracting the $1/\epsilon$ poles we find
$$
\left({1\over \epsilon}{\Gamma(1/2)\over\Gamma(s)}\pi^{s-1/2}+2{1\over
\epsilon}{\zeta(0)\over\Gamma(s)}\pi^{s}\right)\prod_{i<2s+1}e^{-{1 \over\sqrt{2}}\vec\phi\cdot\vec\mu^{i}}
+{\cal O}(\epsilon^0)\eqno(4.4)
$$
which vanishes since $\zeta(0)=-1/2$,  $\Gamma(1/2)=\sqrt{\pi}$. Thus $\Phi_p$ is finite in the limit
$\epsilon\to 0$. The effect of this is to remove the two divergent terms from $\Phi_{p}$ and replace them by
$$
-{ 2\pi^s\over\Gamma(s)}\left({1\over\sqrt{2}} \vec\phi\cdot(\vec\mu^{2s}-\vec\mu^{2s+1}) -(\gamma-{\rm
ln}(4\pi))\right)\prod_{i<2s+1}e^{-{1\over\sqrt{2}}\vec\phi\cdot\vec \mu^{i}}\ ,\eqno(4.5)
$$
which come from the ${\cal O}(\epsilon^0)$ part of the
regularlisation. Here $\gamma$ is the Euler constant. Note that the
effect of this is that the two terms, which are both proportional to
$e^{-{1\over\sqrt{2}} \phi\cdot(\vec \mu^1+\ldots \vec\mu^{2s})}$,
change from having a divergent coefficient to a finite one that
depends linearly on $\vec\phi\cdot(\vec\mu^{2s}-\vec\mu^{2s+1})$.

 When $s=N/2$ there is a $\zeta(1)$ term at $k=N$ but no $\Gamma(0)$ term. Thus we only find the first
contribution to the pole. However we see that, since $\sum_{k=1}^N\vec\mu^k=\vec 0$, the residue of the
$1/\epsilon$ pole is  a constant. In particular we find
$$
{1\over \epsilon}{\pi^{N/2 }\over\Gamma(N/2)} -\sqrt{2}{\pi^{N/2}\over \Gamma(N/2)}
\vec\phi\cdot\vec\mu^N+{\pi^{N/2}\over \Gamma(N/2)}(\gamma-2{\rm ln} (2)-\Gamma'(N/2)/\Gamma(N/2))+{\cal
O}(\epsilon)\ .\eqno(4.6)
$$
Thus we can obtain a well-defined automorphic form by taking
$$
\Phi_{s=N/2} =\lim_{\epsilon\to 0} \left(\Phi_{N/2+\epsilon} -{1\over \epsilon}{\pi^{N/2
}\over\Gamma(N/2)}-{\pi^{N/2}\over \Gamma(N/2)}(\gamma-2{\rm ln}(2)-\Gamma'(N/2)/\Gamma(N/2))\right)\eqno(4.7)
$$
Note that we are also free to remove an arbitrary finite constant from $\Phi_{s+\epsilon}$. Thus in this case
the effect of this is to remove the final, divergent,  term from $\Phi_p$ and replace it with
$$
-\sqrt{2}{\pi^{N/2}\over\Gamma(N/2)} \vec\phi\cdot\vec\mu^N\ .\eqno(4.8)
$$

There is another interesting case where $s=-n$  for some $n=0,1,2,...$. Here the $\Gamma(s)$ term in the
denominator diverges so
$$
\Phi_{s=-n} =0\ .\eqno(4.9)
$$
However the $k=1$ term in $\Phi_p$ survives when $s=0$ (recall that $\zeta(2s)\Gamma(s)$ is
finite when $s$ is a negative integer) and we therefore find
$$
\Phi_{s=0} = -1\eqno(4.10)
$$
for any representation.

\bigskip
{\large {5. Elementary Examples}}
\bigskip
To illustrate the general formalism developed above we now work it out for the simplest cases, namely $SL(2)$
and $SL(3)$.

\bigskip
{\bf {5.1 $SL(2)$}}
\bigskip

The simplest example of an automorphic form is  for the group $SL(2)$ with generators
$$
E_{\beta_1},H, F_{\beta_1}\ ,\eqno(5.1)
$$
with a single positive root generator $E_{\beta_1}$ and a single negative root generator
$F_{\beta_1}=E_{-\beta_1}$ where $\beta_1=\sqrt 2$. The fundamental weight dual to $\beta_1$ is just
$\mu^1=1/\sqrt 2$. We take the local subgroup to be the Cartan involution invariant subgroup which is generated
by $E_{\beta_1}- F_{\beta_1}$. As a result we may choose  the coset representative to be given by
$$
g = e^{\chi E_{\beta_1}}e^{-{1\over \sqrt 2} \phi H }\ ,\eqno(5.2)
$$
The 2-dimensional representation of $SL(2)$ has highest weight $\mu^1 $ and lowest weight $\mu^2 = -\mu^1$.
Taking the states to be normalised so that $E |\mu^2>=|\mu^1>$, we find the action of the Lie algebra elements
on the representation is given by
$$
E_{\beta_1} = \left(\matrix{0&1\cr 0&0\cr}\right)\qquad H  = {1\over \sqrt {2}}\left(\matrix{1&0\cr
0&-1\cr}\right)\qquad F_{\beta_1} = \left(\matrix{0&0\cr 1&0\cr}\right)\ .\eqno(5.3)
$$
Similarly acting with $L(g)$ on the states of the representation we find that
$$\eqalign{
L(g)(m_1|\mu^1>+m_2|\mu^2>) &= (e^{-{\phi/2}}m_1 +m_2e^{{\phi/2}}\chi )|\mu^1> +m_2 e^{{\phi/2}} |\mu^2> \cr
L(g^{-1})(m_1|\mu^1>+m_2|\mu^2>) &= (e^{{\phi/2}}m_1 -m_2e^{{\phi/2}}\chi )|\mu^1> +m_2 e^{-{\phi/2}} |\mu^2> \
. }\eqno(5.4)
$$
Thus acting on the $m_1$ and $m_2$ as a column vector we find that $L (g)$ and $L(g^{-1})$ are represented by
the familiar matrices
$$
e^{\phi/2 }\left(\matrix{e^{-\phi}&\chi \cr 0& 1\cr}\right) \quad e^{\phi/2 }\left(\matrix{1&-\chi \cr 0&
e^{-\phi}\cr}\right) \ .
$$
respectively. One can explicitly compute the non-linear transformation by taking a constant element
$$
g_0 = \left(\matrix{a&b\cr c &d\cr}\right) \in SL(2)\ ,\eqno(5.5)
$$
one then computes $g_0g$ and finds a matrix
$$
h =  \left(\matrix{\cos\theta &-\sin\theta \cr \sin\theta &\cos\theta\cr}\right) \in SO(2)\ ,\eqno(5.6)
$$
such that $g_0g(\phi,\chi)h^{-1} = g(\phi',\chi')$ is again of the form (5.4) but with $\phi\to\phi'$ and
$\chi\to\chi'$.  If we introduce the complex field $\tau = -\chi+ie^{-\phi}$ then we find
$$
e^{i\theta} = {c\tau+d\over |c\tau+d|}\ , \eqno(5.7)
$$
and recover the familiar fractional linear transformation
$$
\tau' = {a\tau+b\over c\tau+d}\ .\eqno(5.8)
$$

As explained in section 2, to construct the automorphic form we first compute the state $|\varphi>$ of equation
(2.9):
$$
|\varphi> = L(g^{-1})( m_1|\mu^1>+ m_2|\mu^2>) =(m_1e^{\phi/2 }-m_2\chi e^{\phi/2 })|\mu^1>+ m_2e^{-\phi/2
}|\mu^2>\ ,\eqno(5.9)
$$
Using equation (2.18) we find the automorphic form to be given by
$$
\Phi_s = \sum_{m_1,m_2\ne (0,0)} {1\over [(m_1-\chi m_2)^2e^{\phi} +m_2^2e^{-\phi}]^s}\ .\eqno(5.10)
$$
Using equation (3.19) we find that the perturbative part is given by
[19]
$$
\Phi_{p} = 2\zeta(2s)e^{-s\phi} + 2\zeta(2s-1)\pi^{1\over 2}{\Gamma(s-{1\over 2})\over \Gamma(s)}e^{(s-1)\phi}\
.\eqno(5.11)
$$
We note that in this simple case $u_1 = m_2^2e^{-\phi}$  and $\tilde
\chi_1 = \chi m_2$ so that, using the equation (3.21) the
non-perturbative part is given by [19]
$$
\Phi_{np} = 4{\pi^s\over \Gamma(s)}e^{-\phi/2}\sum_{\hat m_1=1}^\infty \sum_{m_2\ne 0} \left( \hat m_1 \over
{|m_2|}\right)^{s-{1\over 2}}\cos(2\pi \hat m_1 m_2 \chi)K_{s-{1\over 2}}(2\pi\hat m_1|m_2|e^{-\phi})\
.\eqno(5.12)
$$

We can also consider the 3-dimensional representation of $SL(2)$. The root string in this case is
$$
\{\vec \mu^1,\vec\mu^2,\vec\mu^3\}=   \{\sqrt{2},0,-\sqrt{2}\}\ ,\eqno(5.13)
$$
{\it i.e. } the highest weight is $\vec\mu^1=\sqrt{2}$. In particular we find
$$
\tilde \chi_{12}=\tilde \chi_{23} =\chi\qquad \tilde \chi_{13} =
-{1\over 2}\chi^2\ .\eqno(5.14)
$$
Calculating the non-linearly realised state of equation (2.9) we find that
$$\eqalign{
|\varphi> &= L(g^{-1}) (m_1|\vec\mu^1>+ m_2|\vec\mu^2>
+m_3|\vec\mu^3>) \cr &=(m_1-m_2\chi +{1\over
2}m_3\chi^2)e^{\phi}|\vec\mu^1>+( m_2-m_3\chi)|\vec\mu^2> +
m_3e^{-\phi}|\vec\mu^3>)\ .  }\eqno(5.15)
$$
Then the automorphic form of equation (2.19) becomes
$$
\Phi_s = \sum_{m_1,m_2,m_3\ne (0,0,0)} {1\over [(m_1-\chi
m_2+{1\over2}\chi^2m_3)^2e^{2\phi}+(m_2-\chi
m_3)^2+m_3^2e^{-2\phi}]^s}\ .\eqno(5.16)
$$
Using equation (3.19) the perturbative part is given by
$$
\Phi_{p} = 2\zeta(2s)e^{-2s\phi}  + 2\zeta(2s-1)\pi^{1\over 2}{\Gamma(s-{1\over 2})\over \Gamma(s)}e^{-\phi}
+2\pi {\Gamma(s-1)\over \Gamma(s)}\zeta(2s-2)e^{(2s-3)\phi}\ .\eqno(5.17)
$$
and equation (3.21) the non-perturbative piece by
$$
\eqalign{\Phi_{np} &=  4{\pi^s\over \Gamma(s)}e^{-(s+1/2)\phi}\sum_{\hat m_1=1}^\infty\sum_{(m_2,m_3)\ne (0,0)}
\left( { \hat m_1\over \sqrt{u_1}} \right)^{s-{1\over 2}}\cos(2\pi \hat m_1 \tilde \chi_1)K_{s-{1\over
2}}(2\pi\hat m_1e^{-\phi}\sqrt{u_1})\cr  & +4{\pi^s\over \Gamma(s)}e^{-\phi}\sum_{\hat
m_2=1}^\infty\sum_{m_3\ne 0} \left( {\hat m_2 \over \sqrt{u_2}}\right) ^{s-1}\cos(2\pi \hat m_2 \tilde\chi_2)
K_{s-1}(2\pi\hat m_2\sqrt{u_2})\cr }\ .\eqno(5.18)
$$
Here we note that  $\tilde\chi_1=\chi m_2+\chi^2m_3$, $\tilde\chi_2 = \chi m_3$,  $u_1 = (m_2-\chi
m_3)^2+m_3^2e^{-2\phi}$ and $u_2=m_3^2e^{-2\phi}$. Thus these terms are exponentially suppressed as $
e^{\phi}\to \infty$.

\bigskip
{\bf {5.2 $SL(3)$}}
\bigskip

Let us now consider the 3-dimensional representation  of $SL(3)$. The simple roots are
$$
\vec\alpha_1 =(\sqrt{2},0)\qquad \vec\alpha_2 = (-{1\over\sqrt{2}},\sqrt{{3\over 2}})\ , \eqno(5.19)
$$
and hence the fundamental weights are
$$
\vec\lambda^1 =({1\over \sqrt{2}},{1\over \sqrt{6}})\qquad \vec\lambda^2 = (0,\sqrt{{2\over 3}})\ .\eqno(5.20)
$$
The 3-dimensional representation has highest  weight $\vec\mu^1=\vec\lambda^1$ and the root string is
$$
\vec\mu^1=({1\over \sqrt{2}},{1\over \sqrt{6}})\qquad
\vec\mu^2=\vec\mu^1-\vec\alpha_1=(-{1 \over \sqrt{2}},{1\over
\sqrt{6}}) \qquad \vec\mu^3
=\vec\mu^1-\vec\alpha_1-\vec\alpha_2=(0,-\sqrt{2\over 3})\
.\eqno(5.21)
$$
The fields $\xi$ consist of  $\vec\phi = (\phi,\rho)$  associated to the Cartan subalgebra and
$\chi_{\vec\alpha_{1}}$, $\chi_{\vec\alpha_{1}}$ and $\chi_{\vec \alpha_{1}+\vec\alpha_2}$ are the three
axion-like modes associated to the raising operators $E_{\vec\alpha_1}$,$E_{\vec\alpha_2}$ and
$E_{\vec\alpha_1+\vec\alpha_2}$. We now find
$$\eqalign{
|\varphi> &= L(g^{-1})(m_1|\vec\mu^1 > +m_2|\vec\mu^2 >+
m_3|\vec\mu^3 >)\cr &=  (m_1-m_2\chi_{\vec\alpha_{1}}
-m_3(\chi_{\vec\alpha_1+\vec\alpha_2}-{1\over 2}
\chi_{\vec\alpha_{1}}\chi_{\vec\alpha_{2}}))e^{\phi/2+\rho/2\sqrt
{3}}|\vec\mu^1> \cr
 &+ (m_2-m_3\chi_{\vec\alpha_{2}})e^{-\phi/2+\rho/2\sqrt{3}} |\vec \mu^2 >+
m_3e^{-\rho/\sqrt{3}}|\vec\mu^3 >)\ . \cr} \eqno(5.22)
$$
Following (2.18) we find that the automorphic form is given by
$$
\Phi_s  = \sum_{m_1,m_2,m_3\ne (0,0,0)}{1\over <\varphi|\varphi>^s}\ .\eqno(5.23)
$$
Using equation (3.19) the perturbative part is given by
$$
\eqalign{\Phi_{p} &= 2\zeta(2s)e^{-\sqrt{2}s\vec\phi\cdot\vec\mu^3}+2\zeta(2s-1)\pi^{1 \over
2}{\Gamma(s-{1\over 2})\over \Gamma(s)}e^{-\sqrt{2}\vec\phi\cdot ((s-1/2)\vec\mu^2+1/2\vec\mu^3)}\cr
&\qquad+2\pi {\Gamma(s-1)\over \Gamma(s)}\zeta(2s-2)e^{-\sqrt{2}\vec\phi\cdot((s-1)\vec\mu^1+1/2\vec
\mu^2+1/2\vec\mu^3)}\cr &= 2\zeta(2s)e^{-s\phi}e^{-s\rho/\sqrt{3}} + 2\zeta(2s-1)\pi^{1\over
2}{\Gamma(s-{1\over 2})\over \Gamma(s)}e^{(s-1)\phi}e^{-s\rho/\sqrt{3}}\cr &\qquad +2\pi {\Gamma (s-1)\over
\Gamma(s)}\zeta(2s-2)e^{(2s-3)\rho/\sqrt{3}}}\ . \eqno(5.24)$$ While equation (3.21) gives  the
non-perturbative part to be
$$
\eqalign{\Phi_{np} &= 4{\pi^s\over \Gamma(s)}e^{-{1\over2}(s+1/2)(\phi +\rho/\sqrt{3})}\sum_{\hat
m_1=1}^\infty\sum_{(m_2,m_3)\ne (0,0)} \left( { \hat m_1\over \sqrt {u_1}} \right)^{s-{1\over 2}}\cr
&\hskip2cm\times\cos(2\pi \hat m_1 \tilde \chi_1)K_{s-{1\over 2}}(2\pi \hat
m_1\sqrt{u_1}e^{-{1\over2}(\phi+\rho/\sqrt{3})})\cr  & +4{\pi^s\over
\Gamma(s)}e^{{1\over2}(s-1)\phi-{1\over2}(s+1)\rho/\sqrt {3})}\sum_{\hat m_2=1}^\infty\sum_{m_3\ne 0} \left(
{\hat m_2 \over \sqrt{u_2}}\right) ^{s-1}\cr  &\hskip2cm\times\cos(2\pi \hat m_2 \tilde\chi_2) K_{s-1}(2\pi\hat
m_2\sqrt{u_2}e^{{1\over2}(\phi- \rho/\sqrt{3})})\cr }\  ,\eqno(5.25)
$$
where
$$
u_1 =(m_2-\chi_{\vec\alpha_2}m_3)^2e^{-\phi+\rho/\sqrt{3}} + m_3^2e^ {-2\rho/\sqrt{3}}\ ,\qquad u_2 = m_3^2
e^{-2\rho/\sqrt{3}}\ ,\eqno(5.26)
$$
and
$$
\tilde\chi_1=\chi_{\vec\alpha_1}m_2+(\chi_{\vec\alpha_1+\vec\alpha_2}-{1\over
2} \chi_{\vec\alpha_1}\chi_{\vec\alpha_2})m_3 \qquad \tilde\chi_2 =
\chi_{\vec\alpha_2}m_3\ .\eqno(5.27)
$$

\bigskip
{\large { 6. Perturbative contribution for $E_{n+1}$ Automorphic Forms in terms of $SL(2)\otimes SL(n)$}}
\bigskip

The main purpose for studying the automorphic forms in this paper is to find suitable examples that could occur
in the higher derivative corrections for String Theory compactified to $d=10-n$ dimensions on an
$n$-dimensional torus. Examining these automorphic forms in detail should give detailed information on the
quantum string corrections. In section three, in equation (3.18), we found the ``perturbative" contribution
from the automorphic form for any group $G$ and any representation. The result is expressed in terms of weights
of the representation and in order  to compare to string theory we must express it in terms of string variables
such the string coupling relevant to that dimension and other quantities such as the volume of the compactified
space. In this section we give a more explicit form for the perturbative part by taking explicit expressions
for the weights involving quantities that can then be related to string quantities. This final step and the
comparison to string theory will be given in section seven.

The $E_{n+1}$  symmetry that arises in the compactification of the IIB theory on an $n$ torus has the Dynkin
diagram of figure 1

\bigskip
$$
\matrix{ & & & & & & & &\bullet&\vec\alpha_{n+1} & \cr & & & & & & & & |& &\cr & & & & & & &
&\bullet&\vec\alpha_n&\cr & & & & & & & &|& & \cr \bullet&-&\ldots&-& \bullet&-&\bullet&-&\bullet&-&\bullet&
\cr\vec\alpha_{1}& && &\vec\alpha_{n-4}& &\vec\alpha_{n-3}& & \vec \alpha_{n-2}& &\vec\alpha_{n-1}\cr}
$$
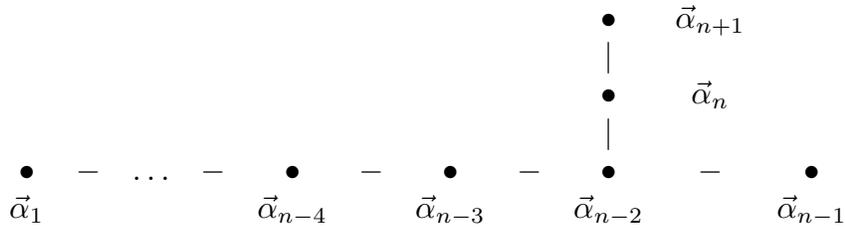
\centerline{FIGURE 1}
\bigskip

The node labeled $n+1$ corresponds to the $SL(2)$  symmetry of the IIB theory while the nodes labeled 1 to
$n-1$ arise are associated with the  part of the  gravity of the IIB theory compactified on the $n$ torus. A
natural decomposition of the $E_{n+1}$ symmetry is found by deleting the node labeled $n$ and analysing the
representations of $E_{n+1}$ that we need in terms of the resulting $SL(2)\otimes SL(n)$ algebra. This deletion
is shown in figure 2.

\bigskip
$$
\matrix{ & & & & & & & &\bullet&\vec\alpha_{n+1} &\cr & & & & & & & & |& &\cr & & & & & & & &\times&&\cr & & &
& & & & &|& &\cr \bullet&-&\ldots&-&\bullet&-&\bullet&-&\bullet&-& \bullet&\cr \vec\alpha_{1}& &&
&\vec\alpha_{n-4}& &\vec\alpha_{n-3}& & \vec\alpha_{n-2}& &\vec \alpha_ {n-1}\cr}
$$
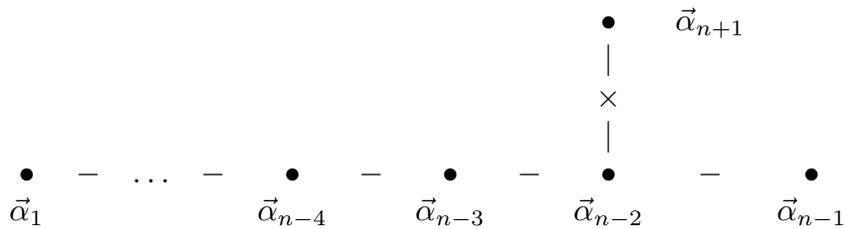
\centerline{FIGURE 2}
\bigskip

This decomposition is very natural when the IIB theory  is
considered from the perspective of $E_{11}$ [39] and [40,41] contain
explicit decompositions of the relevant representations. In addition
a  review on U-duality discussing   $E_{n+1}$ representations is
[42]. Deleting node 10 of the $E_{11}$ Dynkin diagram leads to the
$SL(10)$ content of the ten dimensional $E_{11}$ theory and at low
levels this is just the content of the IIB supergravity theory [43].
The theory in $d$ dimensions arises from the $E_{11}$ theory by
deleting node $d$ of the $E_{11}$ Dynkin diagram leaving the Dynkin
diagrams corresponding to  the $E_{n+1}$ symmetry and $SL(d)$
corresponding to the gravity in the remaining spacetime.

The representations of interest to us are the fundamental
representation associated with node $n+1$. In this section we will
use the above decomposition to find the explicit form of the weights
of this representation in terms of those of  $SL(2) \otimes SL(n)$
and $\vec \phi=(\phi, \rho, \underline \phi )$. Here $\phi$ is
associated to the Cartan subalgebra of $SL(2)$, $\underline\phi$ to
the Cartan subalgebra of $SL(n)$ and $\rho$ is associated with the
Cartan generator of $E_{n+1}$ associated with node $n$ and so does
not appear in $SL(2) \otimes SL(n)$. In section seven we will relate
these fields to the string variables. By substituting these into the
formulae we will obtain the perturbative contribution in a form that
allows its terms to be compared with string theory in $d$
dimensions.

Although we will not use any $E_{11}$ input we will carry out the
decomposition using the techniques employed in the analysis of this
algebra [39-44].  As a first step we write the simple roots of
$E_{n+1}$ as
$$ \vec
\alpha_{n+1}= (\beta_1, 0,0), \qquad \vec  \alpha_n=(0,x,\underline 0)-\vec \nu,\qquad \vec \alpha_i=
(0,0,\underline \alpha_i), \eqno(6.1)
$$
where $\vec \nu= (\mu_1,0,\underline 0)+(0,0,\underline \lambda^{n-2}) $ and $ i=1,\ldots , n-1$. In these
equations $\underline \alpha_i$  and $\underline\lambda^i$  are the simple roots  and fundamental weights of
$SL(n)$ and $\beta_1=\sqrt 2$ and $\mu_1={1\over \sqrt 2}$ the simple root and fundamental weight of $SL(2)$.
Demanding that $\vec \alpha_n^2=2$ we find that
$$
x=\sqrt{{8-n\over 2n}}\ .\eqno(6.2)
$$

We will also need the fundamental weights of $E_{n+1}$, denoted $ \vec\lambda^a, a=1,\ldots , n+1$. These are
readily determined to be
$$
\vec \lambda^i=\left( 0,{1\over x}\underline \lambda^{n-2}\cdot \underline \lambda^i ,\underline
\lambda^i\right),\qquad \vec \lambda^{n}=\left( 0,{1\over x}, \underline 0\right),\qquad \vec
\lambda^{n+1}=\left( \mu_1,{1\over 2 x},\underline 0\right).\ \eqno(6.3)
$$
The reader may verify that $\vec \alpha_a\cdot \vec \lambda^b=\delta_ {a}^b$.

Any root of $E_{n+1}$ can be written as
$$
\vec  \alpha=n_c \vec \alpha_n +m\vec \beta +\sum_i n_i\vec \alpha_i= n_c (0,x,0) -\vec  \lambda \eqno(6.4)
$$
where $ \vec  \lambda= n_c\vec  \nu-\sum_i n_i (0,0,\underline
\alpha_i)-m (\beta_1,0,\underline 0)$ is a weight of $SL(2)\otimes
SL(n)$. If a representation of $SL(2)\otimes SL(n)$ occurs in the
decomposition of the adjoint representation of $E_{n+1}$ its highest
weight must occur for some positive integers $m$,  $n_i$ and $n_c$.
We refer to the integer $n_c$ as the level and we can analyse the
occurrence of highest weights level by level using the techniques of
references [44-46].  Clearly, at level zero i.e $n_c=0$ we have just
the adjoint representation of $SL(2)\otimes SL(n)$. The result is
that the adjoint representation of $E_{n+1}$ contains the adjoint
representation of $SL(2)\otimes SL(n)$ at $n_c=0$ together with the
following highest weight representations of $SL(2)\otimes SL(n)$,
for $n\le 7$,
$$
\matrix {n_c=1 & n_c=2 & n_c=3 & n_c=4 \cr (\mu_1,\underline \lambda^ {n-2})&  (0,\underline \lambda^{n-4})&
(\mu_1,\underline\lambda^{n-6})&  (0,\underline \lambda^{n-1})&\cr} \ . \eqno(6.5)
$$
The weights in the adjoint representation of $E_{n+1}$ are therefore given by
$$\eqalign{
([\beta_1], 0,\underline 0)&\ ,  (0, 0,[\underline \alpha_1+\ldots + \underline \alpha_{n_1}])\ , \
([\mu_1],x,[\underline\lambda^{n-2}])\ ,\ \cr &  (0,2x, [\underline \lambda^{n-4}])\ ,
([\mu_1],3x,[\underline\lambda^{n-6}])\ ,  (0,4x, [ \underline \lambda^{n-1}])} \eqno(6.6)
$$
The first two are simply the weights of the adjoint representation of $SL(2)\otimes SL(n)$. Here $[\bullet]$
denotes the weights in the root string of the $SL(2)$ or $SL(n)$ representation with highest weight $\bullet$.
The reader may verify that one does indeed get the correct number of states  for the adjoint representation of
$E_{n+1}$ for $n=3,\ldots 7$.

We will need to decompose representations of $E_{n+1}$ other than
the adjoint representation. To do this we use the technique of
references [43,47]. If one wants to consider the representation of
$E_{n+1}$ associated with highest weight $\vec\lambda^a$ we add a
node to the $E_{n+1}$ Dynkin diagram which is connected to the node
$a$ by a single line, thereby constructing  the Dynkin diagram for
an enlarged algebra of rank $n+2$. We will denote this additional
node by $\star$ and introduce a corresponding level $n_\star$.
Deleting this node we recover the $E_{n+1}$ Dynkin diagram. The
commutation relations of this new rank $n+2$ algebra respect the
level so the commutation relation between $E_{n+1}$ generators
(level zero) and level one gives level one. As such the level one
generators form a representation of $E_{n+1}$; it is in fact the
representation with highest weight $\vec \lambda^a$. Thus we find
the decomposition of the $\vec\lambda^a$ representation of $E_{n+1}$
into representations of $SL(2)\otimes SL(n)$ by decomposing the
adjoint representation of the enlarged algebra, keeping only
contributions with level $n_\star=1$ and dropping the additional
root $\star$.

Alternatively, but this is a more time consuming construction, one can simply start with the highest weight
$\vec \mu^1$ and lower it with simple roots to construct the entire root string. Given any weight $\vec \mu^i$
in the root string and simple root $\vec\alpha$, one finds that $\vec \mu^i-\vec\alpha$ is also in the root
string provided that $\vec\mu^i\cdot\vec\alpha >0$. In particular the elements of the root string are of the
form $\vec\mu^i = \vec\mu^1- \sum n_{\vec\alpha}\vec\alpha$ where $n_ {\vec\alpha}$ are non-negative integers.

We are most interested in the fundamental representation with
highest weight $\vec\lambda^{n+1} $ of $E_{n+1}$ as this is the
representation from which we construct the automorphic forms that
are relevant to string theory [48]. The weights $\vec \lambda$
appearing in this representation can be written in the form
$$
[\vec\lambda^{n+1}]= \left([\mu], {1\over 2x} -n_c x,[\underline \lambda ]\right)\ . \eqno(6.7)
$$
Here $(\mu,\underline\lambda )$ is the highest weight of the $SL(2)
\otimes SL(n)$ representation. We define an ordering of the weights
as follows, $\vec\mu_i>\vec\mu_j$ if the first non-zero component of
$\vec\mu_i-\vec\mu_j$ in the ordered basis $(0,1,\underline 0) $,
$(1,0,\underline 0) $ and $(0,0,\underline \mu) $ is positive. In
particular the weights are  ordered in terms of increasing level
$n_c$ and then with respect to their $SL(2)$ weights and finally
with respect to their $SL(n)$ weights. This ordering coincides with
that found directly by using the positivity of the roots of the
$E_{n+1}$ algebra.

One finds that the weights in the $\vec\lambda^{n+1} $ representation of $E_{n+1}$ are given by
$$\eqalign{
& ([\mu_1], {1\over 2x},\underline 0)\ ,(0, {1\over
2x}-x,[\underline \lambda^{n-2}])\ ,([\mu_1], {1\over
2x}-2x,[\underline\lambda^{n-4}])\ ,\cr & (0, {1\over
2x}-3x,[\underline\lambda^{n-1}]+[\underline\lambda^{n-5}])\ , (0,
{1 \over 2x}-3x,[\underline\lambda^{n-6}])\ , ([\beta_1], {1\over
2x}-3x,[\underline\lambda^{n-6}])\ ,\cr & ([\mu_1], {1\over
2x}-4x,[\underline\lambda^{n-1}+\underline\lambda^{n-7}])\ ,
([\mu_1], {1\over
2x}-4x,[\underline\lambda^{n-6}+\underline\lambda^{n-2}])\ , \cr
&(0, {1\over
2x}-5x,[\underline\lambda^{n-4}+\underline\lambda^{n-6}])\ ,(0,
{1\over 2x}-5x,[\underline\lambda^{n-1}+\underline\lambda^{n-2}
+\underline \lambda^{n-7}])\ ,\cr &  ([\beta_1], {1\over
2x}-5x,[\underline\lambda^{n-3}+\underline\lambda^{n-7}]) \ ,
([\mu_1], {1\over
2x}-6x,[\underline\lambda^{n-5}+\underline\lambda^{n-7}])\ ,  \cr &
([\mu_1], {1\over 2x}-6x,[\underline 2\lambda^{n-6}])\ , ([\mu_1],
{1\over 2x}-6x,[\underline\lambda^{n-1}+ \underline\lambda^{n-4}
+\underline\lambda^{n-7}])\ ,\ldots }\eqno(6.8)
$$
For small values of $n$ many contributions vanish as one has too
many anti-symmetrised indices. In particular the  $\underline
\lambda^{n-j}$ representation corresponds to totally anti-symmetric
tensors $T^{i_1\ldots i_j}$. The reader may like to verify that one
has the correct count of states for the $5$,$10$,$27$, and
$133$-dimensional representations of $SL(5)$, $SO(5,5)$, $E_6$ and
$E_7$ respectively. To find the 3875 dimensional representation of
$E_8$ one must go further in the analysis.

To continue we label the contributions appearing in  (6.8) by
$\alpha= (n_c,i)$ where the index $i$ labels the different $SL(n)$
representations  at the same level $n_c$. Note that we will
explicitly write out each separate state in the $SL(2)$
representation but not each state in the $SL(n) $ representation. We
choose our labels so that they increase in accord with the ordering
of the weights introduced above. We denote the number of states in
the block labeled $\alpha$ by $d_\alpha$ and define $a_\alpha=\sum_
{\beta<\alpha} d_\beta$ as well as $b_\alpha=\sum_{\beta<\alpha}
n_c(\beta) d_\beta$.
The blocks and their labels are explicitly given below for $SL(n+2)$, $SO(5,5)$,
$E_6$ and $E_7$ in tables 1,2,3,4 respectively.

Substituting the above  weights into the general formula for the perturbative contribution of equation (3.19)
and using that $\vec \phi= (\phi,\rho,\underline \phi )$ we find that this contribution can be written as
$$
\Phi_p= \sum_{n_c,i}\sum_{k=a_{\alpha}+1}^{a_{\alpha+1}} E_k
N_\alpha (\phi)e^{-{s\over\sqrt{2}x}\rho}
e^{(2n_cs-n_ca_\alpha+b_\alpha) {x\over\sqrt{2}} \rho} P_k
(\underline\lambda, \underline \phi) \ ,\eqno(6.9)
$$
where
$$
N_\alpha= 1\ ,\eqno(6.10)
$$
if $\alpha$ corresponds to the singlet representation of $SL(2)$,
$$
N_\alpha=\left\{\matrix{ e^{-\phi(s-{a_\alpha \over 2})} & {\rm\  if\ \ weight}\ \mu_1 \cr
e^{\phi(s-{a_{\alpha+1} \over 2})} & {\rm\  if\   weight\  } -\mu_1 \cr}\right. \ ,\eqno(6.11)
$$
for the doublet representation of $SL(2)$ and
$$
N_\alpha=\left\{\matrix{e^{-\phi(2s-{a_\alpha})}\ \ {\rm\ if\
weight}\  2\mu_1\cr e^{-\phi d_{\alpha-1} }\ \ {\rm\ if\ weight}\ 0
\cr e^{\phi(2s-{a_{\alpha+1}})}\ \ {\rm\  if\   weight}\ -2\mu_1\
\cr}\right. \ ,\eqno(6.12)
$$
for the triplet representation of $SL(2)$. Furthermore we defined
$$
E_k= 2\pi^{{k-1\over 2}}\zeta (2s-k+1)\Gamma (s-{k-1\over 2})/\Gamma (s)\ , \eqno(6.13)
$$
and
$$
P_k(\underline\lambda, \underline \phi)= e^{-\sqrt 2( (s-{k-1\over
2})[\underline \lambda]_{k-a_{\alpha }}+ {1\over 2}([\underline
\lambda]_{1}+\ldots +[\underline \lambda]_{k-1-a_{\alpha}}))\cdot
\underline \phi} \ ,\eqno(6.14)
$$
where $[\underline \lambda]_{r}$ is the $r$-th weight in the root string with highest $SL(n)$ weight
$\underline \lambda$.

\bigskip
{\bf {6.1 $SL(m+2)\to SL(2)\times SL(m)$}}
\bigskip

Let us start by considering the case of $SL(m+2)\to SL(2)\times
SL(m)$ which provides a simple example of the methods we explained
above. The Dynkin diagram of $SL(m+2)$ before and after deleting the
$m$-th node is given in Figure 3.

\bigskip
$$
\matrix{\bullet&\vec\alpha_{m+1} & & & &  &\cr |& &   & & & &\cr
\bullet&\vec\alpha_m&& & &  &\cr |&
 & & && &\cr \bullet&-&\ldots&-&\bullet&&\cr \vec\alpha_{1}& && & \vec\alpha_{m-1}&
 \cr}\qquad\Rightarrow\qquad
\matrix{ \bullet&\vec\alpha_{m+1} && & &  &\cr  |& &   & &&
&\cr\times & & &  &&&\cr |&
 & & && &\cr \bullet&-&\ldots&-&\bullet&&\cr \vec\alpha_{1}& && & \vec\alpha_{m-1}& \cr}
$$
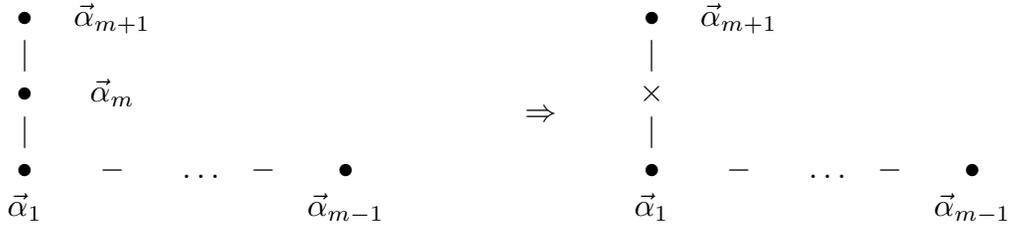
\bigskip \centerline{FIGURE 3}
\bigskip

For $m>1$ the simple roots can be written as
$$
\vec\alpha_{m+1} = (\sqrt{2},0,\underline 0)\qquad \vec\alpha_m =
(-{1\over \sqrt{2}},x,-\underline \lambda^1)\qquad \vec\alpha_{i} =
(0,0,\underline \alpha_i)\ ,\eqno(6.15)
$$
where $x= \sqrt{(m+2)/2m}$ and the fundamental weights are
$$
\vec\lambda^{m+1}= ({1\over \sqrt{2}}, {1\over 2x},\underline
0)\qquad \vec\lambda^m = (0,{1\over x},\underline 0)\qquad
\vec\lambda^{i} = (0,{1\over
x}\underline\lambda^i\cdot\lambda^1,\underline \lambda^i)\
.\eqno(6.16)
$$
For $m=1$, {\it i.e.} $SL(3)$, we have the roots $\vec
\alpha_{m+1}=(\sqrt 2, 0)$ and $\vec \alpha_{m}=(-{1\over \sqrt 2},
x)$ and weights $\vec\lambda^{m+1}=({1\over\sqrt{2}},{1\over 2x})$,
$\vec\lambda^{m}=(0,{1\over x})$.

It is not difficult to see that the weights of the
$\vec\lambda^{m+1}$ representation are
$$
([\mu^1],{1\over 2x},\underline 0),(0,{1\over 2x}- x,[\underline \lambda^1])\ ,\eqno(6.17)
$$
where the first entry is the $SL(2)$ weight and the last the $SL(m)$
weight. Thus we see that $\bf {m+2}\to {(\bf {2},\bf {1})}_{0}\oplus
{(\bf {1},\bf {m})}_1$, where the subscript indicates the level. To
compute the perturbative part of the automorphic form we can now
simply use the formula (6.9). The various parameters needed that are
computed from the group decomposition are listed in Table 1.

\bigskip
 {\offinterlineskip \tabskip=0pt \halign{ \vrule height2.75ex depth1.25ex width 0.6pt #\tabskip=1em
& \hfil #\hfil &\vrule \hfil #\hfil & \hfil #\hfil &\vrule #  &
\hfil #\hfil&\vrule # &  \hfil #\hfil&\vrule # &\hfil #\hfil &
#\vrule width 0.6pt \tabskip=0pt\cr \noalign{\hrule height 0.6pt} &
\omit &&\omit (0,1) && \omit (0,2) &&\omit (1,1) &&\omit   & \cr
\noalign{\hrule}  & $SL(m)$ rep. && ${\underline 0}$ && ${\underline
0}$ && ${\underline \lambda^{1}}$ && &\cr & $SL(2)$ weight && $\mu$
&& $-\mu$ && $0$ &&  &\cr\noalign{\hrule} & $d_\alpha$ && 1 && 1 &&
$n$ && &\cr & $a_\alpha$ && 0 && 1 && 2 && $n+2$ &\cr & $b_\alpha$
&& 0 && 0 && 0 && $n $ &\cr \noalign{\hrule height 0.6pt} }}
\bigskip
{Table 1: $SL(m+2) \to SL(2)\oplus SL(m)$}
\bigskip

Reading off the data from Table 1 and using the formula (6.9) we find
$$
\Phi_p = e^{-s\phi}e^{-{s\over\sqrt{2}x}\rho}E_1 +   e^{(s-1)\phi}e^{-{s\over\sqrt{2}x}\rho}E_2 +
e^{-{s\over\sqrt{2}x}\rho}e^{(2s-2){x\over \sqrt{2}}\rho}\sum_{k=3}^{m+2} E_kP_k(\underline\lambda^1,\underline
\phi)\ ,\eqno(6.18)
$$
where $x=\sqrt{(m+2)/2m}$.

For $m=1$ we find the $\bf 3$ of $SL(3)$ which arises in String Theory as a special case of $SL(3)\times
SL(2)$. In addition for $m=3$ we find the $\bf 5$ of $SL(5)$ which also arises in String Theory. We note that
taking $n=2$ in equation (6.2) gives $x=\sqrt{3/2}$ which agrees with $x=\sqrt{(m+2)/2m}$ for $m=1$. Indeed one
can see that figure 3 agrees with figures 1 and 2 for these two cases, namely $n=2$ and $n=3$, although for
$n=2$ one finds an addition node in figures 1 and 2 corresponding to the extra $SL(2)$ in $SL(3)\times SL(2)$.
For $m\ne 1,3$ we find more general possibilities than those which arise  in string theory.

\vfil

\bigskip
{\bf {6.2 $n=4, E_{5}=SO(5,5)$}}
\bigskip

\bigskip
 {\offinterlineskip \tabskip=0pt \halign{ \vrule height2.75ex depth1.25ex width 0.6pt #\tabskip=1em
& \hfil #\hfil &\vrule \hfil #\hfil & \hfil #\hfil &\vrule #  &
\hfil #\hfil &\vrule #  & \hfil #\hfil &\vrule #  &  \hfil
#\hfil&\vrule # &  \hfil #\hfil&\vrule # &\hfil #\hfil & #\vrule
width 0.6pt \tabskip=0pt\cr \noalign{\hrule height 0.6pt} & \omit
&&\omit (0,1) && \omit (0,2) &&\omit (1,1) &&\omit (2,1) &&\omit
(2,2) &&\omit   & \cr \noalign{\hrule} &  $SL(4)$ rep. &&
${\underline 0}$ && ${\underline 0}$  && ${\underline \lambda^2}$ &&
${\underline 0}$  &&${\underline 0}$  && &\cr & $SL(2)$ weight &&
$\mu$ && $-\mu$ && $0$ && $\mu$ &&$-\mu$&&&\cr \noalign{\hrule} &
$d_\alpha$ && 1 && 1 &&  6 && 1 &&1 &&  &\cr & $a_\alpha$ && 0 && 1
&& 2 && 8 &&9 && 10&\cr & $b_\alpha$ && 0 && 0 && 0 && 6 &&8&&10&\cr
\noalign{\hrule height 0.6pt} }}
\bigskip
{Table 2: $SO(5,5) \to SL(2)\oplus SL(4)$}
\bigskip

$$\eqalign{
\Phi_p &=e^{-s\phi}e^{-{s\over \sqrt{2}x}\rho}E_1 +
e^{(s-1)\phi}e^{-{s\over \sqrt{2}x}\rho}E_2 + e^{-{s\over
\sqrt{2}x}\rho}e^{(2s-2){x\over \sqrt{2}}\rho}\sum_{k=3}^{8}
E_kP_k(\underline\lambda^2,\underline \phi) \cr &+
e^{-(s-4)\phi}e^{-{s\over \sqrt{2}x}\rho}e^{(4s-10){x\over
\sqrt{2}}\rho}E_9P_9(\underline0,\underline \phi)+
e^{(s-5)\phi}e^{-{s\over \sqrt{2}x}\rho}e^{(4s-10){x\over
\sqrt{2}}\rho}E_{10}P_{10}(\underline0,\underline \phi) \
.\cr}\eqno(6.19)
$$
Substituting $x = 1/\sqrt{2}$ gives
$$
\eqalign{\Phi_p&=e^{-s\phi}e^{-s\rho}E_1 +
e^{(s-1)\phi}e^{-s\rho}E_2 + e^{-\rho}\sum_{k=3}^{8}
E_kP_k(\underline\lambda^2,\underline \phi) \cr &+
e^{-(s-4)\phi}e^{(s-5)\rho}E_9P_9(\underline0,\underline \phi)+
e^{(s-5)\phi}e^{(s-5)\rho}E_{10}P_{10}(\underline0,\underline \phi)\
. \cr}\eqno(6.20)
$$

\bigskip
{\bf {6.3 $n=5, E_{6}$}}
\bigskip

\bigskip
 {\offinterlineskip \tabskip=0pt \halign{ \vrule height2.75ex depth1.25ex width 0.6pt #\tabskip=1em
& \hfil #\hfil &\vrule \hfil #\hfil  & \hfil #\hfil &\vrule \hfil
#\hfil & \hfil #\hfil &\vrule # & \hfil #\hfil &\vrule #  & \hfil
#\hfil &\vrule # & \hfil #\hfil&\vrule # &  \hfil #\hfil&\vrule #
&\hfil #\hfil & #\vrule width 0.6pt \tabskip=0pt\cr \noalign{\hrule
height 0.6pt} & \omit &&\omit (0,1) && \omit (0,2) &&\omit (1,0)
&&\omit (2,1) &&\omit (2,2)&&\omit (3,1) &&\omit & \cr
\noalign{\hrule} & $SL(5)$ rep. && ${\underline 0}$ && ${\underline
0}$  && ${\underline \lambda}^3$ && ${\underline \lambda}^1$
&&${\underline \lambda}^1$ && ${\underline \lambda}^4$ && &\cr &
$SL(2)$ weight && $\mu$ && $-\mu$ && $0$ && $\mu$ &&$-\mu$&& $0$&&
&\cr \noalign{\hrule} & $d_\alpha$ && 1 && 1 && 10 && 5 &&5 && 5&&
&\cr & $a_\alpha$ && 0 && 1 && 2 && 12 &&17&&22&& 27 &\cr &
$b_\alpha$ && 0 && 0 && 0 && 10 &&20&&30&& 45 &\cr \noalign{\hrule
height 0.6pt} }}
\bigskip
{Table 3: $E_6 \to SL(2)\oplus SL(5)$}
\bigskip

$$\eqalign{
\Phi_p&= e^{-s\phi}e^{-{s\over \sqrt{2}x}\rho}E_1 +
e^{(s-1)\phi}e^{-{s\over \sqrt{2}x}\rho}E_2 \cr &+ e^{-{s\over
\sqrt{2}x}\rho}e^{(2s-2){x\over\sqrt{2}}\rho}\sum_{k=3}^{12}
E_kP_k(\underline\lambda^3,\underline \phi) +
e^{-(s-6)\phi}e^{-{s\over
\sqrt{2}x}\rho}e^{(4s-14){x\over\sqrt{2}}\rho}\sum_{k=13}^{17}E_kP_k(\underline\lambda^1,\underline
\phi)\cr &+ e^{(s-11)\phi}e^{-{s\over
\sqrt{2}x}\rho}e^{(4s-14){x\over\sqrt{2}}\rho}\sum_{k=18}^{22}E_kP_{k}(\underline\lambda^1,\underline
\phi)+ e^{-{s\over
\sqrt{2}x}\rho}e^{(6s-36){x\over\sqrt{2}}\rho}\sum_{k=23}^{27}E_{k}P_{k}(\underline\lambda^4,\underline
\phi)\ .\cr }\eqno(6.21)
$$
Substituting $x=\sqrt{3/10}$ gives
$$
\eqalign{ \Phi_s&= e^{-s\phi}e^{-{s\sqrt{5\over3}}\rho}E_1 +
e^{(s-1)\phi}e^{-{s\sqrt{5\over3}}\rho}E_2 \cr &+
e^{-(2s+3)\rho/\sqrt{15}}\sum_{k=3}^{12}
E_kP_k(\underline\lambda^3,\underline \phi) +
e^{-(s-6)\phi}e^{(s-21)\rho/\sqrt{15}}\sum_{k=13}^{17}E_kP_k(\underline\lambda^1,\underline
\phi)\cr &+
e^{(s-11)\phi}e^{-(s-6)\phi}e^{(s-21)\rho/\sqrt{15}}\sum_{k=18}^{22}E_kP_{k}(\underline\lambda^1,\underline
\phi)+
e^{(4s-54)\rho/\sqrt{15}}\sum_{k=23}^{27}E_{k}P_{k}(\underline\lambda^4,\underline
\phi)\ .\cr}\eqno(6.22)
$$

\bigskip
{\bf {6.4 $n=6, E_7$}}
\bigskip

\bigskip
 {\offinterlineskip \tabskip=0pt \halign{ \vrule height2.75ex depth1.25ex width 0.6pt #\tabskip=1em
& \hfil #\hfil &\vrule \hfil #\hfil  & \hfil #\hfil &\vrule \hfil #\hfil & \hfil #\hfil &\vrule # & \hfil
#\hfil &\vrule #  & \hfil #\hfil &\vrule # & \hfil #\hfil &\vrule # &  \hfil #\hfil&\vrule # &\hfil #\hfil &
#\vrule  \tabskip=0pt\cr \noalign{\hrule height 0.6pt} & \omit &&\omit (0,1) && \omit (0,2) &&\omit (1,0)
&&\omit (2,1) &&\omit (2,2)&&\omit (3,0)&&\omit (3,1)& \cr \noalign{\hrule} & $SL(6)$ rep.  && ${\underline 0}$
&& ${\underline 0}$ && ${\underline \lambda}^4$ && ${\underline \lambda}^2$ &&${\underline \lambda}^2$ &&
${\underline 0}$ &&  $\underline\theta_\ge$&\cr & $SL(2)$ weight && $\mu$ && $-\mu$ && $0$ && $\mu$ &&$-\mu$&&
$2\mu$ && $0$&\cr\noalign{\hrule} & $d_\alpha$ && 1 && 1 && 15 && 15 &&15 && 1&& 20&\cr & $a_\alpha$ && 0 && 1
&& 2 && 17 &&32&&47&& 48 &\cr & $b_\alpha$ && 0 && 0 && 0 && 15 &&45&&75&& 78 &\cr \noalign{\hrule height
0.6pt} }}
\bigskip

{\offinterlineskip\tabskip=0pt  \halign{ \vrule height2.75ex
depth1.25ex #\tabskip=1em & \hfil #\hfil &\vrule \hfil #\hfil  &
\hfil #\hfil &\vrule \hfil #\hfil & \hfil #\hfil &\vrule # & \hfil
#\hfil &\vrule # & \hfil #\hfil&\vrule # & \hfil #\hfil&\vrule #  &
\hfil #\hfil &\vrule # & \hfil #\hfil &\vrule # & \hfil
#\hfil&\vrule # &\hfil #\hfil & #\vrule
 width 0.6pt \tabskip=0pt\cr
\noalign{\hrule height 0.6pt} & \omit (3,2)&&\omit (3,3) && \omit (3,4) && \omit (3,5) &&\omit (4,1) &&\omit
(4,2) &&\omit (5,1)&&\omit (6,1)&&\omit (6,2)&&  & \cr \noalign{\hrule} & ${\underline 0}$ && ${\underline 0}$
&& $\underline\theta_<$ && ${\underline 0}$ && ${\underline \lambda}^4$ && ${\underline \lambda}^4$ &&
${\underline \lambda}^2$&&${\underline 0}$ && ${\underline 0}$ && &\cr & 0 && 0 && 0&& $-2\mu$ && $\mu$ &&
$-\mu$ && $0$&&$\mu$&& $-\mu$&& &\cr \noalign{\hrule} & 1 && 1 && 15&& 1 && 15 && 15 && 15&&1 && 1&& &\cr & 68
&&69&&70&&85&& 86 && 101 &&116&&131&&132&& 133 &\cr & 138 && 141&&144 && 189 && 192 &&252&&312&& 387&& 393&&
399 &\cr \noalign{\hrule height 0.6pt} }
\bigskip
{Table 3: $E_7 \to SL(2)\oplus SL(6)$. Note that $\underline \theta={\underline \lambda}^1+{\underline
\lambda}^5$ is the highest weight of the adjoint representation of $SL(6)$, $\underline \theta_\ge  $ denotes
the non-negative roots of $SL(6)$ and $\underline \theta_<$ the negative roots.}}
\bigskip

$$\eqalign{
\Phi_p &= e^{-s\phi}e^{-{s\over \sqrt{2}x}\rho}E_1 +
e^{(s-1)\phi}e^{-{s\over \sqrt{2}x}\rho}E_2 \cr &+ e^{-{s\over
\sqrt{2}x}\rho}e^{(2s-2){x\over \sqrt{2}}\rho}\sum_{k=3}^{17}
E_kP_k(\underline\lambda^4,\underline \phi) +
e^{-(s-17/2)\phi}e^{-{s\over \sqrt{2}x}\rho}e^{(4s-19){x\over
\sqrt{2}}\rho}\sum_{k=18}^{32}E_kP_k(\underline\lambda^2,\underline
\phi)\cr &+ e^{(s-47/2)\phi}e^{-{s\over
\sqrt{2}x}\rho}e^{(4s-19){x\over
\sqrt{2}}\rho}\sum_{k=33}^{47}E_kP_{k}(\underline\lambda^2,\underline
\phi)\cr &+ e^{-(2s-47)\phi}e^{-{s\over
\sqrt{2}x}\rho}e^{(6s-66){x\over
\sqrt{2}}\rho}E_{48}P_{48}(\underline 0,\underline \phi)\cr & +
e^{-\phi}e^{-{s\over \sqrt{2}x}\rho}e^{(6s-66){x\over
\sqrt{2}}\rho}\sum_{k=49}^{68}E_{k}P_{k}(\underline\lambda^1+\underline\lambda^5,\underline
\phi)\cr&+ e^{-\phi}e^{-{s\over \sqrt{2}x}\rho}e^{(6s-66){x\over
\sqrt{2}}\rho}E_{69}e^{-{1\over
\sqrt{2}}\sum_{\underline\alpha>0}\underline\alpha\cdot\underline\phi}+
e^{-\phi}e^{-{s\over \sqrt{2}x}\rho}e^{(6s-66){x\over
\sqrt{2}}\rho}E_{70}e^{-{1\over
\sqrt{2}}\sum_{\underline\alpha>0}\underline\alpha\cdot\underline\phi}\cr
&+e^{-\phi} e^{-{s\over \sqrt{2}x}\rho}e^{(6s-66){x\over
\sqrt{2}}\rho}\sum_{k=71}^{85}E_{k}P_{k}(\underline\lambda^1+\underline\lambda^5,\underline
\phi)\cr &+ e^{(2s-86)\phi}e^{-{s\over
\sqrt{2}x}\rho}e^{(6s-66){x\over
\sqrt{2}}\rho}E_{86}P_{86}(\underline 0,\underline \phi) \cr&+
e^{-(s-86/2)\phi}e^{-{s\over \sqrt{2}x}\rho}e^{(8s-152){x\over
\sqrt{2}}\rho}\sum_{k=87}^{101}E_kP_{k}(\underline\lambda^4,\underline
\phi)\cr &+ e^{(s-116/2)\phi}e^{-{s\over
\sqrt{2}x}\rho}e^{(8s-152){x\over
\sqrt{2}}\rho}\sum_{k=102}^{116}E_{k}P_{k}(\underline\lambda^4,\underline
\phi)\cr&+e^{-{s\over \sqrt{2}x}\rho} e^{(10s-268){x\over
\sqrt{2}}\rho}\sum_{k=117}^{131}E_kP_{k}(\underline\lambda^2,\underline
\phi)\cr&+ e^{-(s-131/2)\phi}e^{-{s\over
\sqrt{2}x}\rho}e^{(12s-399){x\over
\sqrt{2}}\rho}E_{132}P_{132}(\underline 0,\underline \phi)\cr &+
e^{(s-133/2)\phi}e^{-{s\over \sqrt{2}x}\rho}e^{(12s-399){x\over
\sqrt{2}}\rho}E_{133}P_{133}(\underline 0,\underline \phi)\ .\cr}
\eqno(6.23)
$$
Note that some care in using (6.9) is required here as the adjoint
representation of $SL(n)$ appearing at level 3 is split by some
$SL(n)$ singlet states. Substituting $x=\sqrt{1/6}$ gives
$$
\eqalign{\Phi_s&= e^{-s\phi}e^{-{s\sqrt{3}}\rho}E_1 +
e^{(s-1)\phi}e^{-{s\sqrt{3}}\rho}E_2 \cr &+
e^{-(2s+1)\rho/\sqrt{3}}\sum_{k=3}^{17}
E_kP_k(\underline\lambda^4,\underline \phi) +
e^{-(s-17/2)\phi}e^{-(s+19/2)\rho/\sqrt{3}}\sum_{k=18}^{32}E_kP_k(\underline\lambda^2,\underline
\phi)\cr &+
e^{(s-47/2)\phi}e^{-(s+19/2)\rho/\sqrt{3}}\sum_{k=33}^{47}E_kP_{k}(\underline\lambda^2,\underline
\phi)+ e^{-(2s-47)\phi}e^{-11\sqrt{3}\rho}E_{48}P_{48}(\underline
0,\underline \phi)\cr & + e^{-\phi}
e^{-11\sqrt{3}\rho}\sum_{k=49}^{68}E_{k}P_{k}(\underline\lambda^1+\underline\lambda^5,\underline
\phi)+ e^{-\phi}e^{-11\sqrt{3}\rho}(E_{69}+E_{70})e^{-{1\over
\sqrt{2}}\sum_{\underline\alpha>0}\underline\alpha\cdot\underline\phi}\cr
&+e^{-\phi}
e^{-11\sqrt{3}\rho}\sum_{k=71}^{85}E_{k}P_{k}(\underline\lambda^1+\underline\lambda^5,\underline
\phi)+ e^{(2s-86)\phi}e^{-11\sqrt{3}\rho}E_{86}P_{86}(\underline
0,\underline \phi) \cr&+
e^{-(s-86/2)\phi}e^{(s-76)\rho/\sqrt{3}}\sum_{k=87}^{101}E_kP_{k}(\underline\lambda^4,\underline
\phi)+
e^{(s-116/2)\phi}e^{(s-76)\rho/\sqrt{3}}\sum_{k=102}^{116}E_{k}P_{k}(\underline\lambda^4,\underline
\phi)\cr&+
e^{(2s-134)\rho/\sqrt{3}}\sum_{k=117}^{131}E_kP_{k}(\underline\lambda^2,\underline
\phi)+
e^{-(s-131/2)\phi}e^{(3s-399/2)\rho/\sqrt{3}}E_{132}P_{132}(\underline
0,\underline \phi)\cr&+
e^{(s-133/2)\phi}e^{(3s-399/2)\rho/\sqrt{3}}E_{133}P_{133}(\underline
0,\underline \phi)\ .\cr}\eqno(6.24)
$$

\bigskip
{\large {7. Applications to Perturbative String Theory}}
\bigskip

We begin with type IIB supergravity in ten-dimensions [4,5,6],
viewed as the effective action of the type IIB superstring. This
theory contains two scalar fields: the dilaton $\phi$ which controls
the string coupling constant, and a RR axion-like field $\chi$. The
metric and scalar part of the lowest order supergravity effect
action is, in Einstein frame,
$$
S = {1\over {\alpha'}^4}\int d^{10}x \sqrt{-g}(R - {1\over 2}\partial_\mu\phi\partial^\mu\phi  -{1\over
2}e^{2\phi}\partial_\mu\chi\partial^\mu\chi)\ .\eqno(7.1)
$$

 Our first step is to show that these fields can be identified with
the fields in the $SL(2)/SO(2)$ coset representative $g$ in (5.2)
that we used above and fix any normalizations. This justifies our
use of the same symbols for both the supergravity fields and the
coset representative. Let us consider the coset $g$ that we
introduced above in (5.2) and consider the Cartan form
$$
g^{-1}\partial_\mu g = -{1\over \sqrt{2}} \partial_\mu\phi H +e^{\phi}\partial_\mu \chi E_{\beta_1}\
.\eqno(7.2)
$$
Under a coset transformation $g\to g_0gh^{-1}$ we see that $g^{-1}\partial_\mu g\to hg^{-1}\partial_\mu gh^{-1}
+ h\partial_\mu h^{-1} $. Since the second term is in  the Lie algebra of $H=SO(2)$ we see that
$$\eqalign{
{\cal P}_\mu &= {1\over 2}g^{-1}\partial_\mu g +{1\over 2}(g^{-1}\partial_\mu g)^T \cr &=-{1\over \sqrt{2}}
\partial_\mu\phi H +{1\over 2}e^{\phi}\partial_\mu \chi (E_{\beta_1}+F_{-\beta_1})}\ .\eqno(7.3)
$$
transforms as ${\cal P}_\mu \to h{\cal P}_\mu  h^{-1}$ under $SL(2)$. The action can now be written in the
manifestly $SL(2)$ invariant form
$$\eqalign{
S &= {1\over {\alpha'}^4}\int d^{10}x \sqrt{-g}(R - tr({\cal
P}_\mu{\cal P}^\mu ))\cr &={1\over {\alpha'}^4}\int d^{10}x
\sqrt{-g}(R - {1\over 2}\partial_\mu\phi\partial^\mu\phi  -{1\over
2}e^{2\phi}\partial_\mu\chi\partial^\mu\chi)\ .}\eqno(7.4)
$$
In particular this shows that we can identify the scalar $\phi$ that appears in the coset with the
ten-dimensional type IIB dilaton and the RR scalar $\chi$ with the scalar field associated to the positive root
generator of $SL(2)$ (up to a possible sign $\chi\to-\chi$).

Let us now consider type IIB string theory compactified on $T^n$. We use the compactification ansatz
$$
ds_{E}^2 = e^{2\alpha\rho}ds_{d}^2 + e^{2\beta\rho}G_{ij}(dx^i+A^i)(dx^j+A^j)\ , \eqno(7.5)
$$
where $G_{ij}$ is metric on the internal $T^n$ with unit determinant. The parameters $\alpha$ and $\beta$ are
chosen to ensure that the reduced action is in Einstein frame (assuming one starts with the ten-dimensional
Einstein frame) and also that the modulus $\rho$ has a standard kinetic term. This determines
$$
\alpha = {1\over 4}\sqrt{n\over 8-n}\qquad \beta =- {1\over 4}\sqrt{8-n\over n}\ .\eqno(7.6)
$$
Note that $\alpha = {1\over4\sqrt{2}x}$ and $\beta =
-{\sqrt{2}x\over4}$ where $x$ arose in the decomposition of
$E_{n+1}\to SL(2)\times SL(n)$ and was determined in (6.2).

The internal metric $G_{ij} = e_i{}^{\overline k}e_{j}{}^{\overline l}\delta_{\overline k\overline l}$ gives
rise to scalar fields in the dimensionally reduced theory. It can be shown that the internal vielbein
$e_i{}^{\overline k}$ is itself a $SL(n)/SO(n)$ coset representative. It therefore contains $n-1$ scalars
$\underline\phi$ that are associated to the Cartan subalgebra of $SL(n)$ as well as $n(n-1)/2$ scalars
$\chi_{\underline\alpha}$ associated to the positive root generators of $SL(n)$. We have the ten-dimensional
dilaton $\phi$ and the volume modulus $\rho$ as well as other scalar fields such as $\chi$ and components of
the $p$-form gauge fields in the internal dimensions.

Thus we see that type IIB supergravity compactified on an $n$ torus
has an $SL(2)\times SL(n)$ symmetry. In fact one finds precisely the
right scalar fields to parameterize an $E_{n+1}/I(E_{n+1})$ coset
$g(\xi)$, where $I(E_{n+1})$ is the Cartan involution invariant
subgroup and also the maximally compact subgroup of $E_{n+1}$. In
particular the fields associated to the Cartan subalgebra of
$E_{n+1}$ are $\phi$, $\rho$ and $\underline \phi$ and these can
identified with
$$
\vec\phi=(\phi,\rho,\underline\phi)\ ,\eqno(7.7)
$$
as we did in the previous section.  It is a remarkable fact that the
entire effective supergravity theory has a $E_{n+1}$ symmetry at
lowest order in derivatives.

As is well-known, a discrete $E_{n+1}$ U-duality is conjecture to
hold in the full quantum string theory. Therefore the complete low
energy effective action containing higher derivative terms must
posses a discrete $E_{n+1}$ symmetry. The terms in the effective
action are made of powers of the various field strengths, which form
representations of $E_{n+1}$, along with functions of the scalar
fields which must be automorphic forms of $E_{n+1}/I(E_{n+1})$.
(Note that particular care must be taken for  $d/2$-form field
strengths which do not generally form $E_{n+1}$ multiplets without
also including their electromagnetic duals.) In particular such a
term has the form
$$
{\cal L}_{\cal O} = \sqrt{-g}\Phi(g) {\cal O}\ ,\eqno(7.8)
$$
where $\Phi$ is an automorphic form and
$$
 {\cal O}\sim D^{2\delta} R^{l_R/2}({\cal P})^{l_1}({\cal F}_2)^{l_2}({\cal F}_3)^{l_3}({\cal F}_4)^{l_4}\
 . \eqno(7.9)
$$
Here $R$ is the Riemann tensor,  ${\cal P}$ the component of the
$E_{n+1}$ Cartan form that is not in the $I(E_{n+1})$ subalgebra and
${\cal F}_{p+2}= L(g^{-1})F_{p+2}$ where $F_{p+2}=dA_{p+1}$ is a
$(p+2)$-form field strength. Note that ${\cal F}_{p+2}$ is
constructed to transform under  local $I(E_{n+1})$ transformations.
Thus all the fields in $\cal O$ transform in some representation of
$I(E_{n+1})$ and U-duality requires that ${\cal L}_{\cal O}$ is
$I(E_{n+1})$-invariant.

In ten dimensions string perturbation theory is an expansion in $g_s = e^{ \phi}$ and naturally takes place in
the so-called string frame where
$$
S = {1\over {\alpha'}^4}\int d^{10} x \sqrt{-g_S}g_s^{-2}R_S+\ldots \eqno(7.10)
$$
Upon reduction to $d=10-n$ dimensions we find
$$
S = {1\over \alpha'^4}\int d^{d} x \sqrt{-g_S}{V_ng_s^{-2}}R_S+\ldots = {1\over  {\alpha'}^{d-2\over 2}}\int
d^{d} x \sqrt{-g_S}g_d^{-2}R_S+\ldots \eqno(7.11)
$$
where $V_n$ is the volume of $T^n$ (in string frame and string units) and $g_d = {\alpha'}^{n\over
4}g_s/\sqrt{V_n}$ is the effective coupling constant in $d$-dimensions. Using our reduction ansatz (7.5) we
find
$$
V_n = e^{n\phi/4+n\beta\rho}\qquad g_d = e^{{8-n\over 8}\phi-n\beta\rho/2}\ . \eqno(7.12)
$$

Since the dilaton $\phi$ is not invariant under a duality transformation the duality structure of the effective
action is most manifest in Einstein frame which, in d-dimensions, is related to the string frame by
$$
(g_E)_{\mu\nu} = g_d^{-{4\over d-2}}(g_S)_{\mu\nu}\ . \eqno(7.13)
$$
When we rescale to string frame the term (7.8) becomes
$$
{\cal L}_{\cal O} = \sqrt{-g_S}g_d^{{4\Delta-2d\over d-2}}\Phi{\cal O}_S\ , \eqno(7.14)
$$
where $\Delta = \delta+{l_R\over 2}+{l_1\over 2}+2{l_2\over 2}+3{l_3\over 2}+4{l_4\over 2}$ counts  the number
of inverse metrics that contained in ${\cal L}_{\cal O}$ and ${\cal O}_S$ denotes $\cal O$ with variables
transformed to the string frame. The perturbative terms that arise in $\Phi$ need to be consistent with
perturbation theory. This requires that, in d-dimensional string frame, each term must be of the form
$g_d^{2g-2}$ where $g=0,1,2,..$ is the genus. Therefore we require that
$$
g_d^{4\Delta-2d\over d-2}\Phi_p\ ,\eqno(7.15)
$$
only contains terms of the form  $g_d^{2g-2}$, $g=0,1,2,...$, which we can identify as arising from
perturbation theory.

Let us examine the automorphic forms of $E_{n+1}$ constructed from representations whose highest weight is
$\vec\lambda^{n+1}$. The perturbative terms where given in the previous section. To make contact with String
Theory here we are interested in their dependence on $g_d$ and $V_n$. Using (6.3) we see that the first term
($k=1$) in $\Phi_p$ is
$$
2\zeta(2s) e^{-s\vec\phi\cdot\vec\lambda^{n+1}}= 2\zeta(2s)g_d^{-{8s\over 8-n}}\ . \eqno(7.16)
$$
Demanding that this comes from lowest order in perturbation theory gives
$$
-{8s\over 8-n}+{4\Delta-2d\over d-2} = -2\ , \eqno(7.17)
$$
and hence we deduce that
$$
s = (\Delta-1)/2\ . \eqno(7.18)
$$

Before we write the complete perturbative parts of the automorphic forms in terms of $g_d$ and $V_n$ it is
helpful to make the following observation. The formula (3.19) the $k$-th term in the perturbative part involves
the weight $\vec w_k=-\sqrt{2}(s-{k-1\over 2})\vec\mu^k-{1\over \sqrt{2}}(\vec\mu^1+\ldots+\vec\mu^{k-1})$.
Thus that the difference between the weights of any two consecutive terms is
$$\eqalign{
\vec w_{k}-\vec w_{k+1}&= -\sqrt{2}(s-{k-1\over2})\vec\mu^{k}
+\sqrt{2}(s-{k\over 2})\vec\mu^{k+1}+{1\over\sqrt{2}}\vec\mu^{k}\cr
&= -\sqrt{2}(s-k/2)(\vec\mu^{k}-\vec\mu^{k+1})\ .\cr}\eqno(7.19)
$$
In particular  we need to evaluate $e^{(\vec w_{k}-\vec
w_{k+1})\cdot \vec\phi}$. Now
$\vec\mu^{k}-\vec\mu^{k+1}=q^i\vec\alpha_i$ is a positive element of
the root lattice (note that the $q^i$ need not all be positive).
From (6.1) we see that
$$\eqalign{
e^{\sqrt{2}\vec\alpha_{n+1}\cdot \vec\phi} &= g_d^{2} V_n\cr e^{\sqrt{2}\vec\alpha_{n}\cdot \vec\phi}
&=V_n^{-{4\over n}}e^{-\sqrt{2}\underline\lambda^{n-2}\cdot \underline\phi}\cr
 e^{\sqrt{2}\vec\alpha_{j}\cdot
\vec\phi} &= e^{\sqrt{2}\underline\alpha_{j}\cdot \underline\phi}\ . \cr}\eqno(7.20)
$$
Thus the power of $g_d$ only changes as we work our way down the root string if $\vec\mu^{k}-\vec\mu^{k+1}$
contains $\vec\alpha_{n+1}$, {\it e.g.} if the $SL(2)$ weight within a given $SL(2)$ representation is lowered.
In this case we see that the power of $g_d$ changes by  an integer multiple of $2s-k$. In addition we see that
the volume dependence only changes if we either change the $SL(2)$ weight or the level $n_c$, {\it i.e.} it is
constant for any given $SL(n)$ representation.

Let us now write down the perturbative automorphic forms of $E_{n+1}$ constructed from the $\vec\lambda^{n+1}$
representation.  To convert the previous formulae it is useful to observe that
$$
e^\phi = g_d V_n^{1\over
2}\qquad e^{\rho\over\sqrt{2}x} = V_n^{-{1\over 2}}g_d^{{n\over 8-n}}\qquad e^{{x\over\sqrt{2}}\rho}=
g_d^{{1\over 2}}V_n^{-{8-n\over 4n}} \ .\eqno(7.21)
$$
Our results for the explicit expressions for $\Phi_p$ that we gave  in section 6 are listed below. We have
written out the expressions with the expectation that the leading order term is tree-level in string theory and
used the fact that $P_k(\underline 0,\underline\phi)=1$. That is we have written the power of $g_d$ required to
convert the result to string frame in front of the automorphic form and so the result in string frame is the
integral $\int d^dx \sqrt{-g}$ times the expressions given below.

\bigskip
{\bf 7.1 $d=10,  E_1=SL(2)$}
\bigskip
 From equation (5.11) we find the perturbative result to be
 $$ g_d^{s-2}\Phi_p = g_d^{-2}E_1 +  g_d^{2s-3}E_2\ , \eqno(7.22)
$$
This is the ten dimensional IIB string theory and this automorphic
form and its relations to string theory has been much studied
[19-26]. We include it here for completeness and to illustrate the
method we are employing. We see that the automorphic form has all
the features that we would expect. In particular the perturbative
part is a power series expansion in $g_s = e^{\phi}$ corresponding
to a tree ($g=0$) and $g=s-{1\over 2}$ correction and is independent
of $\chi$. Thus if we take $s = (\Delta-1)/2$ then we find
contributions at genus $g=0, \Delta/2-1$. We can check that for $s =
{3\over 2}$, that is $R^4$ the power of $g_s$ required to go to
string frame is $g_s^{-{1\over 2}}$ and this is indeed the factor
multiplying the automorphic form above. In addition the
non-perturbative part contains all the $\chi$-dependence and, indeed
$K_{s-1/2}(2\pi m_2|m_1|e^{-\phi})$ is exponentially suppressed as
$e^{\phi} \to 0$. Expanding the Bessel function of equation (5.12)
we find the non-perturbative contribution is given by [23]
$$
\Phi_{np}= {\pi^s\over \Gamma (s)}  \sum _{p\not=0}  \sum _{ m\not=0} e^{2\pi i(pm\chi+i|pm|g_s^{-1})}
\left\vert {p\over m}\right\vert^s {1\over | p|} \sum_{k=0}^\infty {g_s^k\over (4\pi |mp|)^k}{\Gamma (s+k)\over
\Gamma (k+1)\Gamma (s-k)}\ .\eqno(7.23)
$$
We note the typical non-perturbative behaviour $e^{-{2\pi|pm|\over g_s}}$.

\bigskip
{\bf 7.2 $d=8,  E_3=SL(3)\times SL(2)$}
\bigskip

Using equation (7.21) to convert equation (6.18)  to string theory quantities we find the perturbative part of
the $SL(3)$  automorphic function is given by
$$
g_d^{4s/3-2}\Phi_p = g_d^{-2}E_1 +  g^{2s-3}_dV^{s-1/2}_nE_2 + g^ {2s-3}_dV^{-s+3/2}_n E_3\ .\eqno(7.24)
$$
Since this is  just an $SL(3)$ automorphic form the result applies
to terms when there is no $SL(2)$ automorphic form present, but one
can also use it to  give the $SL(3)$ contribution if the latter is
present. Taking $s = (\Delta-1)/2$ then we find contributions at
genus $g=0, \Delta/ 2-1$.

This automorphic form has been conjectured  to arise in String
Theory as a coefficient of the $R^4$ term in 8 dimensions with
$s=3/2$ [27] where a comparison with string theory results was
carried out. In this case it is divergent. The regularization does
not affect $\Phi_{np} $ or the first two terms in $\Phi_{p}$ however
the third term is divergent and, following the discussion in section
4, we find it is, in the $s\to 3/2$ limit
$$
2\pi {\Gamma(s-1)\over \Gamma(s)}\zeta(2s-2)e^{(2s-3)\rho/\sqrt{3}} \longrightarrow {2\pi\over \epsilon}+
4\pi(\gamma-1) + 4\pi\rho/\sqrt{3}+{\cal O}(\epsilon)\ , \eqno(7.25)
$$
where $\gamma$ is the Euler constant. A
suitable renormalized automorphic form is obtained by subtracting off the ${2\pi\over \epsilon}+4\pi(\gamma-1)$
factor.

This coefficient of the  $D^4 R^4$ terms has been conjectured to be
an automorphic form with $s={5\over 2}$ and some checks with string
theory results have been carried out   [27,28].

The non-perturbative part of the $SL(3)$ can be found from equation (3.21) using the change to physical
variables of equation (7.21). We can write the result as $ \Phi_{np}=\Phi_{np}^{(1)}+\Phi_{np}^{(2)} $ where
$$\eqalign{
\Phi_{np}^{(1)}&= {2\pi^s\over \Gamma (s)}(V_n)^{{s\over 2}-{1\over
4}}(g_d)^{-s-{1\over 2}}\sum _{p\not=0} \sum _{(m_2,m_3)\not=(0,0)}
\left\vert {\nu_2\over p^2}\right\vert^{{1\over 4}-{s\over 2}}
e^{2\pi i p\tilde \chi_1} K_{s-{1\over 2}} (2\pi |p||\nu_2|^{{1\over
2}}g_d^{-1}V_n^{-{1\over 2}})\cr \Phi_{np}^{(2)} &= {2\pi^s\over
\Gamma (s)}(V_n)^{{1\over 2}}(g_d)^{{2s\over 3}-1}\sum _{ p\not=0}
\sum _{m_3, m_3\not=0} \left\vert {m_3\over p^2}\right\vert^{1-s}
e^{2\pi i pm_3 \chi_{\vec\alpha_2}} K_{s-1} (2\pi |p m_3|V_n^{- 2})\
.\cr }\eqno(7.26)
$$
In these equations $\nu_2= (m_2-\chi_{\vec\alpha_2} m_3)^2+m_3^2
V_n^2$ and $\tilde \chi_{ 1}
=m_2\chi_{\vec\alpha_1}+m_3\chi_{\vec\alpha_1+\vec\alpha_2}-{1\over
2}\chi_{\vec\alpha_1}\chi_{\vec\alpha_2}$. This result essentially
agrees with that of reference [27], although not in every detail.

It is instructive to use equation (A.4) to carry out the $g_d\to 0$ expansion of the non-perturbative result.
We find that
$$\eqalign{
\Phi_{np}^{(1)}&= {2\pi^s\over \Gamma (s)}V_n^{{s\over 2}+{1\over 4}}g_d^{-s+{1\over 2}}\sum _{ p\not=0}  \sum
_{(m_2 ,m_3)\not=(0,0)} \left\vert {\nu_2\over p^2}\right\vert^{{1\over 4}-{s\over 2}} {e^{2\pi i p \tilde
\chi_1}\over |p\nu_2^{{1\over 2}}|}e^{-2\pi | p\nu_2 |g_d^{-1}V_n ^{-{1\over 2}}} \cr &\times
 \sum_{k=0}^\infty
\left({V_n^{{1\over 2}}g_d\over 4\pi  |p\nu_2^{{1\over 2}}|}\right)^k {\Gamma (s+k)\over \Gamma (k+1)\Gamma
(s-k)}}\eqno(7.27)
$$
while
$$\eqalign{
\Phi_{np}^{(2)}&= {2\pi^s\over \Gamma (s)}V_n^{{3\over
2}}g_d^{{2s\over 3}-1}\sum _{p\not=0}  \sum _{m_3 \not=0} \left|
{m_3\over p^2}\right|^{1-s} {e^{2\pi i (pm_3 \chi_{\vec\alpha_2}+i
|pm_3|V_n^{-2})}\over \sqrt {|pm_3|}} \cr &\times \sum_{k=0}^\infty
\left({V_n^{ 2}\over 4\pi |pm_3|}\right)^k {\Gamma (s+k-{1\over
2})\over \Gamma (k+1)\Gamma (s-k-{1\over 2})}\ . } \eqno(7.28)
$$
We note that the series of terms in the second term always terminates for half integer $s$ and for  $s={3\over
2}$,  that is for $R^4$,    only the first term survives.   Using equation (7.21) to convert to string
variables we find that the second term for $s={3\over 2}$   can be written as
$$
\eqalign{ \Phi_{np}^{(2)}&=2\pi V_n^{{3\over 2}} \sum_{
p\not=0}^\infty\sum_{ m\not=0}^\infty{1\over | m|} e^{2\pi i (p
m\tilde\chi_{\vec\alpha_2}+i|pm|V_n^{-2})} \cr &= 4\pi V_n^{{3\over
2}} \sum_{ p=1}^\infty\sum_{m=1}^\infty{1\over m}{\rm Re}\left
(e^{2\pi im p{\cal T}}\right) \cr &= -8\pi V_n^{{3\over 2}}
\sum_{\hat p=1}^\infty{\rm Re\ ln} \left( 1-e^{2\pi ip{\cal
T}}\right) \ , \cr}\eqno(7.29)
$$
where we have introduced  ${\cal T}= \chi_{\vec\alpha_2} +
iV_n^{-2}$. This term can be interpreted as due to worldsheet
instantons, {\it i.e.} these are non-perturbative in $1/\alpha'$
[27] and first non-perturbative term for $s=3/2$ can be interpreted
as arising from  $(p,q)$-strings [27].

\bigskip
{\bf 7.3 $d=7,  E_4=SL(5)$}
\bigskip
Using equation (7.21) to convert equation (6.18)  to string theory quantities we find the perturbative part of
the $SL(5)$  automorphic function is given by
$$
\eqalign{ g_d^{8s/5-2}\Phi_p = g_d^{-2}E_1 &+ g_d^{2s-3}V^{s-1/2}_nE_2+g_d^ {2s-3}V^{5/6-s/3}_n
\sum_{k=3}^5E_kP_k(\underline\lambda^1,\underline \phi) \ .\cr}\eqno (7.30)
$$
Even though there are five terms there are only two different powers of $g_d$. For $s = (\Delta-1)/2$ then we
find contributions at genus $g=0,\Delta/2-1$ and so we find a physically acceptable perturbative series for any
$\Delta$. For the two cases of most interest, $s=3/2$ and $s=5/2$, $\Phi$ requires regularization. As discussed
in section 4 for $s=3/2$ the divergences arise in the third and fourth terms but they cancel  and the result is
a term with the same power of $g_d$ and $V_n$ but a logarithmic dependence on $\underline\phi$. For $s=5/2$ the
last term is divergent and can be subtracted off, leaving a term proportional to $\ln (g_dV_n^{-5/6})$. Thus
the overall structure remains relatively unchanged, in particular one still finds contributions from two orders
of perturbation theory. The non-perturbative part can be found from equation (3.21) using equation (7.21) to
convert it to string variables.

\bigskip
{\bf 7.4 $d=6,  E_5=SO(5,5)$}
\bigskip
Using equation (7.21) to convert equation (6.21)  to string theory
quantities we find the perturbative part of the $SO(5,5)$
automorphic function is given by
$$
\eqalign{ g_d^{2s-2}\Phi_p &= g_d^{-2}E_1 + g_d^{2s-3}V^{s-1/2}_nE_2 +
g_d^{2s-3}V^{1/2}_n\sum_{k=3}^{8}E_kP_k(\underline\lambda^2, \underline \phi) \cr & + g^{2s-3}_dV^{9/2-s}_nE_9
+ g^{4s-12}_dE_{10} \ .\cr}\eqno(7.31)
$$
We observe that we have ten terms but only three different powers of $g_d$. Looking at the last terms we find
it is a physically acceptable perturbative series if $s\ge {5\over 2}$ and for $s = (\Delta-1)/2$ then we find
contributions at genus $g=0, \Delta/2-1,\Delta-6$. There are divergences for $s\le 5$ but as for $SL(5)$ these
don't significantly affect the powers of $g_d$ that appear, although for $s=5$ the final term is replaced by a
term proportional to $ln(g_d)$, so that only two orders of perturbation theory arise. Hence we find that for
$s={3\over 2}$ that the automorphic form is not relevant to string theory. In the next section we will consider
a constrained SO(5,5) automorphic form, the ten dimensional vector using to construct it is taken to be null.
This always has an acceptable perturbative series which we will discuss there.

\bigskip
{\bf 7.5 $d=5,  E_6$}
\bigskip
Using equation (7.21) to convert equation (6.21)  to string theory
quantities we find the perturbative part of the $E_6$  automorphic
function is given by

$$\eqalign{
g_d^{8s/3-2}\Phi_p &= g_d^{-2}E_1 + g_d^{2s-3}V^{s-1/2}_nE_2 +
g_d^{2s-3}V^{s/5+3/10}_n\sum_{k=3}^{12}E_kP_k(\underline\lambda^3, \underline \phi) \cr & +
g^{2s-3}_dV^{-3s/5+51/10}_n\sum_{k=13}^{17}E_kP_k(\underline\lambda^1, \underline \phi) +
g^{4s-20}_dV^{2s/5-17/5}_n\sum_{k=18}^{22}E_{k}P_{k}(\underline \lambda^1,\underline \phi)\cr &+
g^{4s-20}_dV^{-2s/5+27/5}_n\sum_{k=23}^{27}E_{k}P_{k}(\underline \lambda^4,\underline \phi)  \ .\cr}\eqno(7.32)
$$
We have twenty seven terms but only three different powers of $g_d$. It is  a physically acceptable
perturbative series only if $s\ge {9\over 2}$. There are divergences if $s\le 27/2$ but these don't
significantly alter the powers of $g_d$ that appear. If we take $s = (\Delta-1)/2$ then we find contributions
at genus $g=0, \Delta/2-1,\Delta-10$. We could consider an automorphic form that obeys an $E_6$-invariant cubic
constraint and like the $SO(5,5)$ case this may well always have an acceptable perturbative series.

\bigskip
{\bf 7.6 $d=4,  E_7$}
\bigskip
Using equation (7.21) to convert equation (6.23)  to string theory
quantities we find the perturbative part of the $E_7$ automorphic
function is given by

$$\eqalign{
g_d^{4s-2}\Phi_p &= g_d^{-2}E_1 +   g_d^{2s-3}V^{s-1/2}_nE_2 \cr &+
g_d^{2s-3}V^{s/3+1/6}_n\sum_{k=3}^{17}
E_kP_k(\underline\lambda^4,\underline \phi) +
g_d^{2s-3}V^{-s/3+35/6}_n\sum_{k=18}^{32}E_kP_k(\underline\lambda^2,
\underline \phi)\cr &+
g_d^{4s-35}V^{2s/3-61/6}_n\sum_{k=33}^{47}E_kP_{k}(\underline
\lambda^2,\underline \phi)+ g^{2s+12}_dV_n^{-s+29}E_{48}\cr&+
g_d^{4s-36}V^5_n
\sum_{k=49}^{68}E_{k}P_{k}(\underline\lambda^1+\underline\lambda^5,\underline
\phi)+g_d^{4s-36}V^5_n(E_{69}+E_{70})e^{-{1\over \sqrt
{2}}\sum_{\underline\alpha>0}\underline\alpha\cdot\underline\phi}\cr
& + g_d^{4s-36}V^5_n
\sum_{k=71}^{85}E_{k}P_{k}(\underline\lambda^1+\underline\lambda^5,\underline
\phi) +g_d^{6s-121}V^{s-75/2}_nE_{86} \cr&+g_d^
{{4s-35}}V^{s/3-53/6}_n\sum_{k=87}^{101}E_kP_{k}(\underline\lambda^4,
\underline \phi)+
g_d^{6s-136}V^{-2s/3+125/3}_n\sum_{k=102}^{116}E_{k}P_{k}
(\underline\lambda^4,\underline
\phi)\cr&+g_d^{6s-136}V^{-s/3+67/3}_n\sum_{k=117}^{131}E_kP_{k}
(\underline\lambda^2,\underline \phi)\cr&+
g_d^{6s-136}V^{1/2}_nE_{132}+ g_d^{8s-268}V^{-s+133/2}_nE_{133}\ .
\cr} \eqno(7.33)
$$
There are 133 terms but, in  contrast to above,  we find quite a few
different powers of the coupling constant $g_d$. We observe that
there is no value of $s$ for which the series is acceptable as it
will for any $s$ involve odd and even powers of $g_d$. However we
expect that the correct automorphic form is likely to be constructed
by imposing a  quartic $E_7$-invariant constraint on the lattice.

\bigskip
{\large {8.  Perturbative Evaluation of the Constrained SO(5,5) Automorphic Form.}}
\bigskip

The vector representation of $SO(5,5)$  takes the from
$$
|\psi >= \sum_{i=1}^{5}n^i |\vec\mu^i>+  \sum_{i=6}^{10}m_i |\vec\mu^i> \eqno(8.1)
$$
Here the $n^i$ and $m_i$ belong to the $\bf \bar 5$ and $\bf 5$ representations of the $SL(5)$ subgroup.  Since
we are dealing with a vector it is $SO (5,5)$ invariant to  impose  that its length vanishes. This corresponds
to the constraint
$$
\sum_{i=1}^{5}n^i \tilde m_i=0 \eqno(8.2)
$$
where $\tilde m_i=m_{11-i}$. As such rather than sum over the
$SO(5,5)$ lattice as before we can sum subject to this constraint.
Such a possibility was considered in reference [29] and we will use
some of the technical tricks used there. In this section we will
evaluate only the perturbative contribution of the corresponding
automorphic form and so  we can  set all $\chi_{\vec\alpha}=0$  from
the outset.  As such, we find that
$$
|\varphi> = L(g^{-1}) |\psi>=\sum_{i=1}^{5} n^i e^{{1\over \sqrt{2}}\vec\phi\cdot\vec \mu^i} |\vec\mu^i> +
\sum_{i=6}^{10}\tilde m_i e^{{1\over \sqrt{2}}\vec\phi\cdot \vec \mu^i}|\vec\mu^i>\ . \eqno(8.3)
$$
Taking $u= <\varphi |\varphi >$ we must evaluate
$$
\Phi(\xi) = \sum_{\Lambda_c} {1\over (u(\xi))^s}=\sum_\Lambda {\pi^s\over\Gamma(s)}\int_0^1 d\theta
\int_0^\infty {dt\over t^{1+s}}e^{-{\pi\over t}u}e^{2\pi i\theta \sum  \tilde m_i n^i  \tilde m_i} \
,\eqno(8.4)
$$
where $\Lambda$ is the sum over all integers in the lattice;  the above constraint being  implemented  by the
integral over $\theta$.  We may write the sum over the ten integers as
$$
   \sum _{\Lambda}=  \sum _{n^i\ ,\ \tilde m_i=0}^\wedge+ \sum_{n^i} \sum_{\tilde m_i}^\wedge \eqno(8.5)
$$
where the hat means that the term with all the integers vanish is excluded. The first sum leads to the
expression
$$
\Phi_{1}\equiv \sum _{n^i}^\wedge {1\over  [(n^1)^2e^{\sqrt 2 \vec\phi\cdot\vec \mu^1}+\ldots +(n^5)^2e^{\sqrt
2 \vec\phi\cdot\vec \mu^5}]^s}\ . \eqno(8.6)
$$
This closely resembles the automorphic form for $SL(5)$ but the weights are not the same since  they are the
first five weights that occur in the ten representation of $SO(5,5)$. Thus the perturbative formula for
$\Phi_1$ simply consists of the first five terms in the perturbative part of the unconstrained $SO(5,5)$
automorphic form.

Let us  denote the second term of equation (8.4) arising from the split in the sum given in equation (8.5) by
$\Phi_2$. It can be evaluated by apply the Poisson resummation formula to to the five integers  $n^i$. This is
possible as the sum of $n^i$ is over all integers. Using equation (A.2),  the result is
$$\eqalign{
\Phi_2&=\sum_{\hat n_1} \sum_m^\wedge {\pi^s\over\Gamma(s)}\int_0^1 d\theta \int_0^\infty {dt\over
t^{1+s-{5\over 2}}} e^{-{1\over \sqrt 2} \vec\phi\cdot(\vec\mu^1+\ldots +\vec\mu^5)}\cr & \hskip2cm
e^{-{\pi\over t}\sum_{i=6}^{10} \tilde m_i^2 e^{\sqrt 2 \vec\phi\cdot\vec \mu^i}} e^{-{\pi t}\sum_{i=1}^{5}
(\hat n_i+\theta \tilde m_i)^2  e^{-\sqrt 2 \vec\phi\cdot\vec \mu^i}} } \eqno(8.7)
$$
We observe that  $\hat n_i\to \hat n_i + \tilde m_i$ has the same effect as taking $\theta \to \theta +1$. As
such we may restrict the sum to $\hat n_i$ modulo $\tilde m_i$,  but take the integral over $\theta$ to be from
$-\infty $ to $\infty$. Completing the square on $\theta$ and changing to the variable $y$ we find the
expression becomes
$$\eqalign{
\Phi_2&=\sum_{\hat n\ {\rm mod }\tilde m} \sum_{\tilde m_i}^\wedge {\pi^s\over\Gamma(s)} \int_0^\infty {dt\over
t^{1+s-{4\over 2}}}\int_{-\infty}^\infty dy e^{-\pi y^2} e^{-{1\over \sqrt 2} \vec\phi\cdot (\vec\mu^1+\ldots
+\vec\mu^5)}\cr &\hskip2cm \times  e^{-{\pi\over t}\sum_{i=6}^{10} \tilde m_i^2  e^{\sqrt 2 \vec\phi\cdot\vec
\mu^i}} e^{-{\pi t}\left(-{( \hat n\diamond  \tilde m )^2\over \tilde m \diamond \tilde m} +\hat n\diamond \hat
n\right)} {1\over \sqrt {\tilde m\diamond \tilde m}}\ ,} \eqno(8.8)
$$
where
$$ p\diamond  q= \sum _{i=1}^5 p_iq_ie^{-\sqrt 2 \vec\phi\cdot\vec
\mu^i}\ , \eqno(8.9)
$$
and
$$
y=(\theta +{\hat n \diamond \tilde m\over \tilde m\diamond  \tilde
m})\ \sqrt{t \tilde m\diamond  \tilde  m }\ . \eqno(8.10)
$$
We may carry out the integral over $y$ which gives the factor $1$. The expression of equation (8.8) can be
broken into two terms depending  if
$$
{(\hat n\diamond   \tilde m)^2 }=(\hat n\diamond \hat n)(  \tilde
m\diamond  \tilde m) \eqno(8.11)
$$
or not. If it does not then we find a Bessel function which does not contain the perturbative terms we are
trying to compute and so we discard this term. By the Schwarz inequality equation (8.11) is only satisfied if
$\hat n_i=\lambda \tilde m_i$ where $\lambda$ is an integer such that this relation holds. The number of
solutions for fixed $ \tilde m_i$ is just the greatest common divisor (gcd)  of $ \tilde m_i$, {\it i.e.} gcd
($ \tilde m_i$). Thus our expression becomes
$$\eqalign{
\Phi_2&= \sum_m^\wedge {\pi^s\over\Gamma(s)} \int_0^\infty {dt\over
t^{1+s-{4\over 2}}} e^{-{1\over \sqrt 2}
\vec\phi\cdot(\vec\mu^1+\ldots +\vec\mu^5)} {e^{-{\pi\over
t}\sum_{i=6}^{10} \tilde m_i^2  e^{\sqrt 2 \vec\phi\cdot\vec \mu^i}}
\over \sqrt { \tilde m\cdot  \tilde m}}gcd ( \tilde m_i)\cr
&=\sum_m^\wedge {\pi^s\over\Gamma(s)}{\Gamma (s-{4\over 2})\over
\pi^{s-{4\over 2}}} e^{-{1\over \sqrt 2}
\vec\phi\cdot(\vec\mu^1+\ldots +\vec\mu^5)} {gcd ( \tilde m_i)\over
( {  \tilde m\diamond  \tilde  m })^{s-{5\over 2}+1}} }\eqno(8.12)
$$

In carrying out this step we have used that
$$
 \tilde m\diamond  \tilde
  m= \sum _{i=1}^5 \tilde m_i^2 e^{-\sqrt 2 \vec\phi\cdot\vec \mu^i} =\sum_{i=6}^{10} \tilde m_i^2
e^{\sqrt 2 \vec\phi\cdot\vec \mu^i}\ , \eqno(8.13)
$$
due to the fact that for the vector representation $\vec\mu^i=-\vec\mu^{11-i}$ and that $\tilde m_i\tilde
=m_{11-i}$.

To process this expression further we make use of the formulae
$$
\sum_r^\wedge {1\over  ( r\cdot r )^l}= \sum _x^\wedge {1\over x^{2l}}\sum_{r^\prime, coprime}^\wedge {1\over (
r^\prime\cdot r^\prime )^l} = \zeta (2l) \sum_{r^\prime, coprime}^\wedge {1\over ( r^\prime\cdot r^\prime )^l}
\ ,\eqno(8.14)
$$
which arises from taking out the gcd $x$ out of $r$. We use this formula to express the sum of $\tilde m_i$ in
terms of coprimes and combine the gcd divisor which emerges together with the one already there and write the
result in terms of a zeta function. We then use the formula again to rewrite the sum of coprimes in terms of an
ordinary sum. The result of all this is that our expression is now given by
$$
\Phi_2=\pi^2 {\Gamma (s-{4\over 2})\over \Gamma (s)}
{\zeta(2s-4)\over \zeta (2s-3)}e^{-{1\over \sqrt 2}
\vec\phi\cdot(\vec\mu^1+\ldots +\vec\mu^5)}\sum_{\tilde m}^\wedge
{1\over (\tilde m\diamond \tilde m)^{s-{3\over 2}}} \ .\eqno(8.15)
$$
Thus we find that the peturbative contribution to $SO(5,5)$ automorphic constructed  using the vector
representation is given by
$$\eqalign{
\Phi(\xi) &= \Phi_1+\Phi_2\cr &=\sum _{n}^\wedge {1\over
[(n^1)^2e^{\sqrt 2 \vec\phi\cdot\vec \mu^1}+\ldots +(n^5)^2e^{\sqrt
2 \vec\phi\cdot\vec \mu^5}]^s} \cr &\hskip 1cm + \pi^2 {\Gamma
(s-{4\over 2})\over \Gamma (s)} {\zeta(2s-4)\over \zeta
(2s-3)}\sum_{\tilde m}^\wedge {e^{-{1\over \sqrt 2}
\vec\phi\cdot(\vec\mu^1+ \ldots +\vec\mu^5)}\over [\tilde m_1^2
e^{-\sqrt 2 \vec\phi\cdot\vec \mu^1}+\ldots +\tilde m_5^2 e^{-\sqrt
2 \vec\phi\cdot\vec \mu^5}]^{s-{3\over 2}} }\ .}\eqno(8.16)
$$

We recognise the first term as having the same form as we found for $SL(5)$ although it is important to
remember that the weights that occur here are those for the vector representation  of $SO(5,5)$, but only for
the $\bf \bar 5$ part in the decomposition to $SL(5)$.  However we can still apply equation (3.19) to find that
the result is given by
$$\eqalign{
\Phi_1&= e^{-s\phi}e^{-{s\rho\over \sqrt 2 x}}(E_1+e^{(2s-1)\phi}E_2
+e^{s\phi} e^{(s-1)2x \rho\over \sqrt 2 }\sum_{k=3}^5 E_k
P_k(\underline\lambda^2,\underline\phi) )\cr &= g_d^{2-2s}
\left(g_d^{-2}E_1+g_d^{(2s-3)}(V_n^{s-{1\over 2}}E_2+V_n^{{1\over
2}} \sum_{k=3}^5 E_k
P_k(\underline\lambda^2,\underline\phi))\right)\ .} \eqno(8.17)
$$
In the last line we have used the formula (7.21) relevant for $SO(5,5)$, that is taken $n=4$ to convert from
$\phi$ and $\rho$ to the physical variables $g_d$ and $V_n$.

We now evaluate the second term. Using the expression for the weights of the vector representation given in
equation (6.3) we find that
$$
e^{-{1\over \sqrt 2} \vec\phi\cdot(\vec\mu^1+\ldots +\vec\mu^5)}=
e^{-{\rho\over \sqrt 2 x}}e^{-{ \sqrt 2}\underline\lambda^1\cdot \phi}=
V_n^{1\over 2}g_d^{-1}e^{-{  \sqrt 2}\underline\lambda^1\cdot \phi}\ .
\eqno(8.18)
$$
Using equation (2.25) we may rewrite $\Phi_2$ as
$$
\Phi_2= \pi^{2s-{7\over 2}} {\Gamma (s-2)\Gamma (4-s)\over \Gamma
(s)\Gamma (s-{3\over 2})} {\zeta(2s-4)\over \zeta
(2s-3)}\sum_m^\wedge {1\over ( m_1^2 e^{\sqrt 2 \vec\phi\cdot\vec
\mu^1}+\ldots +\tilde m_5^2 e^{\sqrt 2 \vec\phi\cdot\vec \mu^5}
)^{4-s}} \ .\eqno(8.19)
$$
We can use our previous formulae for $SL(5)$ to evaluate this term and then convert it to physical variables
using $n=4$ in equation (7.21) to find
$$\eqalign{
\Phi_2&=\pi^{2s-{7\over 2}} {\Gamma (s-2)\Gamma (4-s)\over \Gamma
(s)\Gamma (s-{3\over 2})} {\zeta(2s-4)\over \zeta (2s-3)}
e^{(s-4)\phi}e^{{(s-4)\rho\over \sqrt 2 x}}\cr &\hskip
2cm\times\left(E^\prime_1+e^{(-2s+7)\phi}E^\prime_2 \
+e^{(-s+4)\phi} e^{(-s+3)2x \rho\over \sqrt 2 }\sum_{k=3}^5
E^\prime_k P_k(\underline\lambda^2,\underline\phi)\right) \cr
&=\pi^{2s-{7\over 2}} {\Gamma (s-2)\Gamma (4-s)\over \Gamma
(s)\Gamma (s-{3\over 2})} {\zeta(2s-4)\over \zeta
(2s-3)}g_d^{2s-8}\cr &\hskip2cm
\left(E^\prime_1+g_d^{(-2s+7)}(V_n^{-s+{7\over
2}}E^\prime_2+V_n^{-{1\over 2}} \sum_{k=3}^5 E^\prime_k
P_k(\underline\lambda^2,\underline\phi))\right)} \eqno(8.20)
$$
where $E_k^\prime= 2\pi^{k-1\over 2} \zeta (-2s+9-k) \Gamma
(-s+{9\over 2}-{k\over 2})/\Gamma (-s+4)$.

Let us now consider the case $s={3\over 2}$. We observe that the $
\Phi_2$ part given in equation (8.15) contains an automorphic form
at the value $s-{3\over 2}=0$ which is equal to $-1$. The resulting
contribution from $\Phi_2$ is therefore
$$
{2\over 3} \pi^{ 2}e^{-{1\over \sqrt 2}
\vec\phi\cdot(\vec\mu^1+\ldots +\vec\mu^5)}={2\over 3} \pi^{ 2}
g_d^{-1}V_n^{{1\over 2}}e^{-{ \sqrt 2}
\vec\phi\cdot\underline\lambda^1}\ . \eqno(8.21)
$$
One can find the same result from equation (8.20) where it is the
last term that gives a non-zero result. Hence we find that for
$s={3\over 2}$ which corresponds to $R^4$ in seven dimensions we
find that the result is given in Einstein frame by
$$
\int d^6 x\sqrt{-g} R^4 \Phi_{{3\over 2}}\ , \eqno(8.22)
$$
where the perturbative part is given by
$$
\Phi_{p,{3\over 2}}=g_d^{-3} (E_1+g_d^{2}(V_nE_2+V_n^{{1\over 2}}
\sum_{k=3}^5 E_k P_k(\underline\lambda^2,\underline\phi)))+{2\over
3} \pi^{ 2} g_d^{-1}V_n^{{1\over 2}}e^{-{ \sqrt 2}
\vec\phi\cdot\vec\lambda^1}\ . \eqno(8.23)
$$
We observe that it only has six  terms as opposed to the ten terms
in the unconstrained automorphic form of equation (7.31). However
the first five terms are just the same as the first five terms of
this unconstrained automorphic form. To move to string frame one
requires a factor of $g_d$ and so we find the only contributions are
at tree level and one loop. Hence we note, that unlike the
unconstrained $SO(5,5)$ automorphic form, it gives an acceptable
result.

Let us now consider the case of $s={5\over 2}$. Examining equation
(8.20) we see that the prefactors are finite except for a divergent
numerator factor of  $\zeta(1)$. However, there is also a $\zeta(1)$
divergent factor in the last of the first terms in $\Phi_1$ which
being at the end of the set of terms is not canceled by any term.
Thus we find six terms which are divergent and examining them one
finds that they are contained in an $SO(5,5)$ automorphic form for
$s={3\over 2}$. Indeed we may write
$$
\Phi_{p,{5\over 2}}= {\rm first\ four\ terms\ of\ }\Phi_1 +
4\zeta(1) \Phi_{p,{3\over 2}}\ . \eqno(8.24)
$$
However, we can regulate this in an $SO(5,5)$ manner  by shifting $s\to s+\epsilon$, 
so that $\zeta(1)= \zeta(2s-4) \sim 1/2\epsilon$, and subtracting off and
entire $SO(5,5)$ automorphic form $2\epsilon^{-1}\Phi_{3/2}$ (note that this also requires that the divergent
non-perturbative part also cancels - as it must since  regularization preserves the automorphic property).
These correspond to a $R^6$ term in seven dimensions and and so we find that the effective action contains in
Einstein frame the term
$$
\int d^6 x\sqrt{-g} R^6 \Phi_{{5\over 2}} \ . \eqno(8.25)
$$
where the perturbative part is given by
$$
\Phi_{p,{5\over 2}}= g_d^{-5} (E_1+g_d^{4}(V_nE_2+V_n^{{1\over 2}}
\sum_{k=3}^4 E_k P_k(\underline\lambda^4,\underline\phi)))\ .
\eqno(8.26)
$$
To move to string frame we require a factor of $g_d^3$ and so we
find only a tree and two loop correction. This is in contrast to the
unconstrained form of equation (7.31) which possess ten terms all of
which are acceptable from a perturbative viewpoint and indeed they
contain the same powers of the coupling constant.

For $s\ge {7\over 2}$ we find that all the ten terms in $SO(5,5)$ constrained automorphic form are finite and
they have an acceptable form when viewed from the string perspective. Indeed we find that in string frame the
automorphic form corresponding to $D^{2\delta}R^{{l_R\over 2}}$, that is $s={\delta+l_R-2\over 4}$ has the
following powers of the coupling
$$
g_d^{-2},\ g_d^{{\delta+l_R-8\over 2}},\ g_d^{\delta+l_R-12}\ .
\eqno(8.27)
$$
In fact the  unconstrained automorphic form of equation (7.31) also has an acceptable coupling constant
dependence which is the same as the constrained automorphic form. Furthermore the first five terms are the
same. This strongly suggests that the perturbative parts of the constrained and unconstrained automorphic forms
are the same. If so this would allows us to construct automorphic forms with no perturbative part by taking
their difference.

Finally for the sake of completeness we note that we could have
computed the ``perturbative" part of the $SO(n,n)$ automorphic form
in essentially the same way. In which case  the analogue of equation
(8.17) is given by
$$\eqalign{
\Phi(\xi) &=\sum _{n}^\wedge {1\over  (n_1^2e^{\sqrt 2
\vec\phi\cdot\vec \mu_1}+\ldots +n_n^2e^{\sqrt 2 \vec\phi\cdot\vec
\mu_n})^s} \cr & + \pi^{{n\over 2}-{1\over 2}}{\Gamma (s-{n\over
2}+{1\over 2})\over \Gamma (s)} {\zeta(2s-n+1)\over \zeta
(2s-n+2)}\sum_m^\wedge {e^{-{1\over \sqrt 2}
\vec\phi\cdot(\vec\mu_1+\ldots +\vec\mu_n)}\over (\tilde m_1^2
e^{-\sqrt 2 \vec\phi\cdot\vec \mu_1}+\ldots +\tilde m_n^2 e^{-\sqrt
2 \vec\phi\cdot\vec \mu_n})^{s-{n\over 2}+1}} \ ,}\eqno(8.28)
$$
and the formula (3.19) can be used to extract the perturbative part.

\bigskip
{\large {9. Discussion}}
\bigskip

In this paper we have  used our previous method [31] to  construct
Eisenstein-like automorphic forms for any group and representation
and found  explicit formulae for the ``perturbative" and
``non-perturbative" parts in terms of the weights of the
representation. Applying this to  the groups  $E_{n+1}$  and  the
fundamental representation associated with node   ${n+1}$, we have
explicitly  computed the perturbative part in terms of the  string
coupling $g_d$ in $d$ dimensions and the volume of the torus $V_n$.
We  then examined the result to see if it could be physical, that is
compatible with the result of a string theory calculation. In
particular we looked for to see if it only contains terms of the
form $g_d^{2g-2}$ for $g$ a positive integer. We have taken care to
find expressions whose dependence on string quantities is very
transparent. We note that the derivation is almost entirely group
theoretic in nature involving the manipulation of properties of the
representation and its decomposition into representations of
$SL(2)\otimes SL(n)$; the $SL(2)$ part being the well known symmetry
of the IIB ten dimensional theory and the $SL(n)$ arising as the
manifest symmetry of the $n$-torus.

For the case of dimensions $d\ge 7$, that is automorphic forms of $SL(N)$ groups based on the $\bf {\bar  N}$
representation, we find that this is always the case    and the  perturbative series  has only two terms.   In
six dimensions, we considered  the automorphic forms for the group $SO(5,5)$ based on the vector, {\it i.e.}
the ${\bf 10}$ representation. The perturbation expansion is physical for $s\ge 5/2$, containing contributions
from two ($s=5/2$) or three ($s>5/2$) orders of perturbation theory. However the $s=3/2$ automorphic form that
occurs with the $R^4$ term does not have a good perturbation expansion. To rectify this we also considered the
constrained $SO(5,5)$ automorphic form based on a null vector representation. The resulting series are
physical, containing contributions from two orders of perturbation theory,  and we have computed them
explicitly. 


In five  dimensions we considered the automorphic form for $E_6$
constructed from the $\bf 27$ representation. The resulting
perturbation expansions are physical if $s\ge 9/2$, where it only
contains contributions from two ($s=9/2$) or three ($s>9/2$) orders
in perturbation theory. Since the $\bf 27$ representation of $E_6$
admits a cubic invariant there is a constrained automorphic form for
$E_6$ which may have better agreement with string perturbation
theory. In four dimensions we considered the automorphic form for
$E_7$ constructed from the $\bf 133$ representation and it appears
to be unable to agree with string perturbation theory. In this case
there is a quartic invariant of $E_7$ and therefore one can define a
constrained $E_7$ automorphic form. The perturbative contributions
from such constrained $E_6$ and $E_7$ automorphic forms is currently
under investigation.  In particular it is of interest to obtain
automorphic forms which are consistent with the non-renormalization
theorems of [34].

One important question is what representation should one take to
construct the automorphic forms. In this paper we have taken the
fundamental representations associated with the node $n+1$ of the
$E_{n+1}$ Dynkin diagram (see figure 1). This is supported by the
dimensional reduction calculation of [48], along the lines of
reference [31], and  outlined in the introduction, to identify the
highest weight contained in the automorphic form. However it would
also be of interest to evaluate the perturbative parts of
automorphic forms based on other representations and see if they
could be relevant to String Theory.

The automorphic forms we have considered are only convergent if $s>
{N\over 2}$ where $N$ is the dimensions of the representation. In
ten dimensions, where $N=2$, this condition is met for all terms of
interest. However in compactified String Theory one readily finds
that the automorphic forms that arise as coefficients for higher
derivative terms are naively ill-defined at low orders. In this
paper we used analytic continuation to define the automorphic forms
for more general values of $s$. This still leaves some values of $s$
where a regularization scheme is required due to poles in the
complex $s$ plane. In particular we chose to deform $s\to
s+\epsilon$ and then subtract any poles in $\epsilon$. Since this
procedure preserves the automorphic property of $\Phi$ one sees that
in general the residue of any pole in $\epsilon$ must itself be an
automorphic form. In the simplest examples, such as unconstrained
automorphic forms with $s=N/2$, the residue is just a constant.
However more generally there can be situations where the residue is
itself a non-trivial automorphic form that needs to be subtracted
off, as was the case for the constrained $SO(5,5)$ automorphic form
with $s=5/2$. Therefore one can expect there to be a rich interplay
between regularization and automorphic forms that would be
interesting to explore.


Non-holomorphic automorphic forms are non-analytic, unlike their
better known cousins. However, their behaviour is partly controlled
if they are required to obey Laplace type equations and, as
advocated in [31], similar equations related to all the higher order
Casimirs of the group from which they are constructed. The
unconstrained automorphic forms for $SO(5,5)$ and $E_{n+1}$ for
$n=5,6.7$ are, following arguments given in [29],  unlikely to obey
such equations and one must adopt constraints in the sum over the
integers to recover these equations. It would be interesting to
investigate these equations more systematically for the automorphic
forms considered in this paper. We also note that the non-Eisenstein
automorphic forms found [23] in ten dimensions obey Laplace
equations with sources. Clearly, there is much to be understood
about non-holomorphic automorphic forms.

\bigskip
{\bf Note Added}: 
It is instructive to compare our results with those of [37]. In the case of the fundamental representation of $SL(n)$ our automorphic forms agree with theirs and so do the corresponding perturbative parts. 
For $SO(5,5)$ these authors demonstrate that (their equation (3.54), in our notation) 
$$
\Phi^{SO(5,5)}_{s=3/2} = g_{d}^{-1}\left({2\zeta(3)\over g^2_d}+2\Phi^{SO(4,4)}_{s=1}\right)\ . \eqno(9.1)
$$
If we use equations (3.19) and (8.28), applied to $SO(4,4)$, we find 
$$\eqalign{
\Phi^{SO(4,4)}_{p,s=1} &= 2\zeta(2)e^{-\sqrt{2}\vec\nu^1\cdot\vec\phi} + 2\pi\zeta(1)e^{-{1\over\sqrt{2}}(\vec\nu^2+\vec\nu^1)\cdot\vec\phi} -  \pi\Gamma(0)e^{-{1\over\sqrt{2}}(\vec\nu^2+\vec\nu^1)\cdot\vec\phi}\cr &+
{1\over 3}\pi^2e^{{1\over\sqrt{2}}\vec\nu^4\cdot\vec\phi} e^{-{1\over\sqrt{2}}(\vec\nu^3+\vec\nu^2+\vec\nu^1)\cdot\vec\phi}+{1\over 3}\pi^2 e^{-{1\over\sqrt{2}}(\vec\nu^4+\vec\nu^3+\vec\nu^2+\vec\nu^1)\cdot\vec\phi}
\ ,\cr}
$$
(note that, strickly speaking, we should replace $s=1$ with $s=1+\epsilon$ to obtain a finite answer). Here $\vec\nu^a$, $a=1,2,3,4$ are the first 4 weights of the 8-dimensional representation of $SO(4,4)$. 
The relevent weights to take are $\vec\nu^a=\vec\mu^{a+1}$, where $\vec\mu^i$, $i=1,2,3,4,5$ are the first 5 weights of the 10-dimensional representation of $SO(5,5)$.  
Substituting this into (9.1) and using (7.21) to write $V_n = e^{-\sqrt{2}\vec\nu^1\cdot \vec\phi}$,  one readily sees that the perturbative part of (9.1) is precisely (8.23). 

Thus, for dimensions six and above,  the perturbative parts of the automorphic forms in [37] agree with those found here. Therefore it is natural to expect that the complete automorphic forms used in [37] are equal to the ones we have defined here by anayltic continuation. An exception is the seven dimensional case where an additional automorphic form constructed from the $\bf 10$ of $SL(5)$ appears in [37], which was not considered in this paper.   In addition we have proposed the $s=5/2$ automorphic form defined below (8.24) for $SO(5,5)$, which was not considered in [37]. We showed that it has a good perurbative expansion and, like the $s=3/2$ case, contains far fewer terms than those the occur at generic values of $s$. 
 


\bigskip
{\large {Acknowledgements}}
\bigskip

Peter West would like to thank The Erwin Schroedinger International
Institute for Mathematical Physics and the Theoretical Physics
Department of the Vienna University of technology for their kind
hospitality in October 2009 when some of the work in this paper was
carried out. We would also like to thank Finn Gubay for useful
comments. This work has been supported by an STFC Rolling grant
ST/G000395/1.

\bigskip
{\large {Appendix: Formulae}}
\bigskip

Here we list some formulae that are used through the main text.
$$
{1\over u^s} = {\pi^s\over \Gamma(s)}\int_0^\infty {dt\over t^{1+s}}e^{-{\pi u \over t}}\ .\eqno(A.1)
$$

Poisson resummation formula:
$$
\sum_{\vec m\in {\bf Z}^N} e^{-\pi (\vec m-\vec a)\cdot A(\vec m-\vec a)+2\pi i\vec m\cdot \vec b} =
\sum_{\vec{\hat m}\in {\bf Z}^N} {\rm det A}^{-{1\over 2}} e^{-\pi (\vec{\hat m}+\vec b)\cdot A^{-1}(\vec{\hat
m}+\vec b)+2\pi i(\vec{\hat m}+\vec b)\cdot \vec a}\ .\eqno(A.2)
$$

Bessel function integral identity
$$
\int_0^\infty {dt\over t^{1+\lambda}}e^{-at-b/t} = 2\left|{a\over b}\right|^{\lambda/2}K_\lambda(2\sqrt{|ab|})\
.\eqno(A.3)
$$
Asymptotic behaviour of Bessel function as $x\to \infty$
$$
K_\lambda(z) = \sqrt{\pi\over 2z}e^{-z}\sum_{l=0}^\infty {1\over
(2z)^l}{\Gamma(\lambda+l+{1\over2})\over\Gamma(l+1)\Gamma(\lambda-l+{1\over 2})}\ .\eqno(A.4)
$$

For two fundamental weights of $SL(n)$:
$$
\underline \lambda^i\cdot\underline \lambda^j ={i(n-j)\over n}\qquad i\le j\ .\eqno(A.5)
$$

\bigskip
{\large {References}}
\bigskip

\item{[1]}
     I.~C.~G.~Campbell and P.~C.~West,
     Nucl.\ Phys.\ B {\bf 243}, 112 (1984).

\item{[2]}
     F.~Giani and M.~Pernici,
     Phys.\ Rev.\ D {\bf 30}, 325 (1984).

\item{[3]}  M.~Huq and M.~A.~Namazie,
     Class.\ Quant.\ Grav.\  {\bf 2}, 293 (1985)
     [Erratum-ibid.\  {\bf 2}, 597 (1985)].

\item{[4]}
J.~H.~Schwarz and P.~C.~West,
     Phys.\ Lett.\ B {\bf 126}, 301 (1983).

\item{[5]}
     P.~S.~Howe and P.~C.~West,
     Nucl.\ Phys.\ B {\bf 238}, 181 (1984).

\item{[6]}
     J.~H.~Schwarz,
     Nucl.\ Phys.\ B {\bf 226}, 269 (1983).

\item{[7]}
     E.~Cremmer, B.~Julia and J.~Scherk,
     Phys.\ Lett.\ B {\bf 76}, 409 (1978).

\item{[8]}
E.~Cremmer and B.~Julia,
Phys.\ Lett.\ B {\bf 80}, 48 (1978).

\item{[9]}
N.~Marcus and J.~H.~Schwarz,
     Nucl.\ Phys.\ B {\bf 228} (1983) 145.

\item{[10]}
  B.~Julia,
  Invited talk given at Johns Hopkins Workshop on Current Problems in Particle Theory, Baltimore, Md., May 25-27, 1981.
Published in Johns Hopkins Wkshp.1981:23

\item{[11]}H.~Nicolai,
  Phys.\ Lett.\  B {\bf 194} (1987) 402.

\item{[12]}
B.~Julia and H.~Nicolai,
     Nucl.\ Phys.\ B {\bf 482}, 431 (1996)
     [arXiv:hep-th/9608082].

\item{[13]}
B.~Julia, in {\it Vertex Operators and Mathematical Physics},
Publications of the Mathematical Sciences Research Institute no3.
Springer Verlag (1984); in {\it Superspace and Supergravity}, ed.
S.~W.~Hawking and M.~Rocek, Cambridge University Press (1981)

\item{[14]} C. Teitelboim, Phys. Lett. B167 (1986) 69.

\item{[15]}  R. Nepomechie , Phys. Rev. D31, (1984) 1921;

\item{[16]}A.~Sen,
     Nucl.\ Phys.\ B {\bf 404}, 109 (1993)
     [arXiv:hep-th/9207053].

\item{[17]}
A.~Font, L.~E.~Ibanez, D.~Lust and F.~Quevedo,
     Phys.\ Lett.\ B {\bf 249}, 35 (1990).

\item{[18]}
C.~M.~Hull and P.~K.~Townsend,
     Nucl.\ Phys.\ B {\bf 438}, 109 (1995)
     [arXiv:hep-th/9410167].

\item{[19]}
M.~B.~Green and M.~Gutperle,
     Nucl.\ Phys.\ B {\bf 498}, 195 (1997)
     [arXiv:hep-th/9701093].

\item{[20]}
     M.~B.~Green, M.~Gutperle and P.~Vanhove,
     Phys.\ Lett.\ B {\bf 409} (1997) 177
     [arXiv:hep-th/9706175].

\item{[21]}M.~B.~Green and S.~Sethi,
     Phys.\ Rev.\ D {\bf 59}, 046006 (1999)
     [arXiv:hep-th/9808061].

\item{[22]} M.~B.~Green, H.~h.~Kwon and P.~Vanhove,
     Phys.\ Rev.\ D {\bf 61}, 104010 (2000)
     [arXiv:hep-th/9910055].

\item{[23]}M.~B.~Green and P.~Vanhove,
     JHEP {\bf 0601}, 093 (2006)
     [arXiv:hep-th/0510027].

\item{[24]} M.~B.~Green, J.~G.~Russo and P.~Vanhove,
     arXiv:hep-th/0610299.

\item{[25]}
A.~Basu,
     arXiv:hep-th/0610335.

\item{[26]}
N.~Berkovits and C.~Vafa,
     Nucl.\ Phys.\ B {\bf 533}, 181 (1998)
     [arXiv:hep-th/9803145].

\item{[27]}
E.~Kiritsis and B.~Pioline,
(p,q) string instantons,''
     Nucl.\ Phys.\ B {\bf 508}, 509 (1997)
     [arXiv:hep-th/9707018].

\item{[28]}
A.~Basu,
     arXiv:hep-th/07121252;
  Phys.\ Rev.\  D {\bf 77} (2008) 106004
  [arXiv:0712.1252 [hep-th]].

\item{[29]}
   N.~A.~Obers and B.~Pioline,
   Commun.\ Math.\ Phys.\  {\bf 209}, 275 (2000)
   [arXiv:hep-th/9903113].

\item{[30]}
N.~Lambert and P.~West,
Phys.\ Rev.\ D {\bf 74}, 065002 (2006) [arXiv:hep-th/0603255].

\item{[31]}
N.~Lambert and P.~West,
Phys.\ Rev.\ D {\bf 75}, 066002 (2007) [arXiv:hep-th/0611318].

\item{[32]}
  L.~Bao, M.~Cederwall and B.~E.~W.~Nilsson,
  Class.\ Quant.\ Grav.\  {\bf 25} (2008) 095001
  [arXiv:0706.1183 [hep-th]].

\item{[33]}
L.~Bao, J.~Bielecki, M.~Cederwall, B.~E.~W.~Nilsson and D.~Persson,
  JHEP {\bf 0807} (2008) 048
  [arXiv:0710.4907 [hep-th]].

\item{[34]}
N.~Berkovits,
  Phys.\ Rev.\ Lett.\  {\bf 98} (2007) 211601
  [arXiv:hep-th/0609006].

\item{[35]}
L.~Bao, A.~Kleinschmidt, B.~E.~W.~Nilsson, D.~Persson and B.~Pioline,
  arXiv:0909.4299 [hep-th].

\item{[36]}
 B.~Pioline and D.~Persson,
  arXiv:0902.3274 [hep-th].

\item{[37]}
 M.~B.~Green, J.~G.~Russo and P.~Vanhove,
  arXiv:1001.2535 [hep-th].

\item{[38]}  B.~Pioline,
  arXiv:1001.3647 [hep-th].

\item{[39]}  P.~C.~West,
     Class.\ Quant.\ Grav.\  {\bf 18}, 4443 (2001)
     [arXiv:hep-th/0104081].

\item{[40]}
P. West,
JHEP 0408 (2004) 052, hep-th/0406150.

\item{[41]}
P.~P.~Cook and P.~C.~West,
  JHEP {\bf 0811} (2008) 091
  [arXiv:0805.4451 [hep-th]].

\item{[42]}A.~Obers and B.~Pioline,
  Phys.\ Rept.\  {\bf 318} (1999) 113
  [arXiv:hep-th/9809039].

\item{[43]}
I.~Schnakenburg and P.~C.~West,
   Phys.\ Lett.\ B {\bf 517}, 421 (2001)
   [arXiv:hep-th/0107181].

\item{[44]}
M.R Gaberdiel, D. I. Olive and P. West.
Nucl. Phys. {\bf B 645}
(2002) 403-437, hep-th/0205068.

\item{[45]}  T.~Damour, M.~Henneaux and H.~Nicolai,
     Phys.\ Rev.\ Lett.\  {\bf 89}, 221601 (2002)
     [arXiv:hep-th/0207267].

\item{[46]}
P. West,
Class. Quant. Grav. {\bf 20} (2003) 2393,  hep-th/0307024.

\item {[47]}  A. Kleinschmidt and P. West,
JHEP 0402 (2004) 033,  hep-th/0312247.

\item{[48]}F.~Gubay, N.~Lambert and P.~West,
  arXiv:1002.1068 [hep-th].

\end